\documentclass[twocolumn,trackchanges]{aastex62}
\usepackage{graphicx}
\graphicspath{{./}{figures/}}
\defcitealias{2020MNRAS.498.2757G}{GA20}
\defcitealias{2022MNRAS.514.3285G}{GA22}

\shorttitle{Broadband variability in GX 339-4}
\shortauthors{Tanenia et al.}

\begin{document}

\title{\textbf{Modelling the Energy-dependent broadband variability in the black-hole transient GX 339-4 using \textit{Astrosat} and \textit{NICER}}}

\author{Hitesh Tanenia}
\altaffiliation{tanenia.hitesh@gmail.com}
\affiliation{Department of Physics, Jamia Millia Islamia, Jamia Nagar, New Delhi-110025, India}

\author{Akash Garg}
\affiliation{Inter-University Centre for Astronomy and Astrophysics, Ganeshkhind, Pune-411007, India}

\author{Ranjeev Misra}
\affiliation{Inter-University Centre for Astronomy and Astrophysics, Ganeshkhind, Pune-411007, India}

\author{Somasri Sen}
\affiliation{Department of Physics, Jamia Millia Islamia, Jamia Nagar, New Delhi-110025, India}



\begin{abstract}

We present a spectro-timing analysis of the black hole X-ray transient GX 339-4 using simultaneous observations from \textit{Astrosat} and \textit{NICER} during the 2021 outburst period. The combined spectrum obtained from \textit{NICER, LAXPC, and SXT} data is effectively described by a model comprising a thermal disk component, hard Comptonization component, and reflection component with an {\tt edge}. Our analysis of the \textit{Astrosat} and \textit{NICER} spectra indicates the source to be in a low/hard state, with a photon index of $\sim$ 1.64. The Power Density Spectra (PDS) obtained from both \textit{Astrosat} and \textit{NICER} observations exhibit two prominent broad features at 0.22 Hz and 2.94 Hz. We generated energy-dependent time lag and fractional root mean square (frms) at both frequencies in a broad energy range of 0.5-30 keV and found the presence of hard lags along with a decrease in variability at higher energy levels. Additionally, we discovered that the correlated variations in accretion rate, inner disc radius, coronal heating rate, and the scattering fraction, along with a delay between them, can explain the observed frms and lag spectra for both features.

\end{abstract}



\section{Introduction} \label{sec:intro}
Black Hole X-ray Binaries (BHXB) are classified into two primary categories: transient and persistent. Transient systems predominantly remain in a quiescence (low flux) state. On the other hand, persistent systems consistently maintain an active state throughout their observed lifespan. The majority of BHXBs are considered to be transients and exhibit phases of quiescence along with occasional bright outbursts (high flux) characterized by rapid fluctuations in their spectral and temporal properties \citep{2006ARA&A..44...49R,2011BASI...39..409B}. During the outburst, a BHXB transient typically transits through four different spectral states: the Hard State (HS), the Hard Intermediate State (HIMS), the Soft Intermediate State (SIMS), and the Soft State (SS). The HS is characterized by a dominant hard spectrum with a power law nature. The non-thermal emission produced by a hot and optically thin plasma known as the corona dominates the X-ray spectra of BHXB in the HS \citep{1976ApJ...204..187S}. The Soft State (SS) is characterized by the predominance of soft photons originating from the geometrically thin and optically thick accretion disk \citep{1976MNRAS.175..613S}. Both soft and hard components contribute to the energy spectrum in the intermediate states.

BHXBs exhibit rapid timing variability. The quantification of this rapid variability is achieved through the Power Density Spectrum (PDS). The PDS is obtained by squaring the Fourier amplitude derived from the Fourier transformation of the light curve \citep{2005AIPC..797..345V}. The PDS exhibits broadband noise-like patterns and occasionally narrow peaks called quasi-periodic oscillations (QPOs) \citep{2000MNRAS.318..361N,2002ApJ...572..392B,2006ARA&A..44...49R,2014SSRv..183...43B,2019NewAR..8501524I,Mendez2021}. These QPOs have been categorized as Low-Frequency QPOs (LFQPOs, $\leq$ 30 Hz) and High-Frequency QPOs (HFQPOs,$\geq$ 60 Hz). Only a small number of BHXB have exhibited HFQPOs, which have a central frequency in the range of around 30 Hz to hundreds of Hz (e.g., \cite{2012MNRAS.426.1701B,2013MNRAS.435.2132M,2022MNRAS.517.1469M}). These QPOs are less common than LFQPOs. The PDS is often fitted using multiple Lorentzians or power law components to calculate the characteristic frequency ($\nu$), quality factor\footnote{The quality factor is defined as the ratio of the characteristic frequency to the full width
at half maximum (FWHM).}  (Q=$\nu$/FWHM), and strength of observed features. These QPOs have also been classified as Type A, B, and C based on quality factors (see review by \cite{2019NewAR..8501524I}).

The significant contribution to the X-ray variability in BHXB is due to the aperiodic broadband noise components in the PDS. Such variability doesn't exhibit a smooth distribution; instead, it displays enhanced power around a particular frequency, causing multiple bumps in the PDS. This distinct shape arises from combining multiple broad features, manifesting as peaked noise components within the PDS. \cite{1997MNRAS.292..679L} suggested that mass accretion rate fluctuations in the disc can propagate inwards and generate broadband noise. \cite{2006MNRAS.367..801A} adopted a numerical method to explain the spectral and timing characteristics of X-ray light curves from black hole X-ray binaries (BHXRB) and active galactic nuclei (AGN). Their model suggests that the variability in the X-rays is caused by the fluctuation in the accretion rate, with these fluctuations occurring on the viscous time scale. \cite{2016ApJ...832..181V} proposed a model aimed at providing a possible explanation for the peaked noise that emerged in the PDS. They interpret that such variability is produced by the interference of two X-ray components, disk Comptonization and synchrotron Comptonization, with a time lag between them. \cite{2020MNRAS.496.3808B} performed general relativistic magnetohydrodynamics (GRMHD) simulations to explain the broadband variability observed in BHXB. They suggest the mass-flux variations within the accretion flow could explain such variability. To identify multiple variability features, they took into account the mass accretion rate as a proxy for luminosity. Their results provide evidence for inward propagating fluctuations.

Likewise, several models have been put forward to identify the origin of the QPOs. For instance, \citet{stella1997lense,stella1999correlations} proposed that precession of the innermost part of the disc, which is dominated by the Lense-Thirring effect, is the cause of the peaks seen in the PDS. \cite{2006ApJ...642..420S} developed a precessing ring model to describe the low-frequency QPOs observed in BHXB. The model is based on an inclined ring of hot gas that orbits around the black hole. For Kerr black holes, this ring undergoes precession around the spin axis of the black hole at the Lense-Thirring frequency. They have shown the relationship between the radius of the ring and black hole spin for a specific QPO frequency, providing a lower limit for the spin parameter. \cite{2020ApJ...889L..36M} characterized the QPO frequency as the relativistic dynamical frequency of a truncated accretion disk.

Both QPOs and broadband noise possess energy-dependent properties such as fractional rms (frms) and time lags. Several works have identified the correlation of frms and lags with coronal and disc properties, implying the possibility of the radiative origin of the variability. For instance, \cite{2022MNRAS.514.2891Z} used RXTE \citep[Rossi X-ray Timing Explorer][]{2006ApJS..163..401J} observations to investigate the high-frequency variability for GRS 1915+105. They showed that the frms amplitude of a high-frequency bump depends on the hardness ratio and frequency of the type-C QPO. They found a correlation between the coronal temperature and the frms amplitude of the bump. \cite{2022NatAs...6..577M} suggested a possible link between the emergence of the bump and coronal region in GRS 1915+105. According to \cite{2014A&ARv..22...72U}, at relatively high frequencies, the broadband noise occasionally exhibits soft lags caused by X-ray reverberation of corona photons reflected by the accretion disc, while at low frequencies, it shows hard lags, which are assumed to be originated from the Comptonization.

To understand which radiative mechanism is responsible for the observed variability, different models have been proposed and applied to fit the time lags and frms. For instance, \cite{2014MNRAS.445.2818K} developed a model to explain the energy-dependent flux variability of photons from a thermal Comptonizing plasma oscillating at kilohertz frequencies. They considered the oscillatory fluctuations in the soft photon source or the plasma heating rate. They applied this model on 4U 1608-52, showing that a model with the size of the Comptonizing plasma of $\sim$ 1 km can accurately predict both the observed soft time-lags and the frms. \cite{karpouzas2020comptonizing,bellavita2022vkompth} used the time-dependent Comptonization model VKOMPTHDK, which takes into account thermal reprocessing in the disc and Compton upscattering in the corona, explains both the hard and soft lags, without considering the dynamical origin of the QPOs. Similarly, \cite{2023MNRAS.520..113R} utilized the same Comptonization model to fit the frms and time lag spectra obtained for type-C QPO using \textit{NICER} data for MAXI J1535-571. \cite{2019MNRAS.486.2964M} developed a stochastic propagation model that quantitatively explains temporal behavior. Their model, defined by only three parameters: inner disk radii, truncation temperature, and the time lag between them, successfully fitted the energy-dependent fractional root mean square and time lag observed for Cygnus X-1. Using the same stochastic model, \cite{2019ApJ...887..101J} and \cite{2020ApJ...889L..17M} explained the temporal characteristics of Swift J1658.2–4242 and MAXI J1820+070, respectively, as observed by \textit{Astrosat}.

\cite{2020MNRAS.498.2757G} introduced a generic scheme to interpret time lag and frms. Their model was based on the assumption that variations in the radiative components could reproduce the observed timing features, such as time lag and frms. In their methodology, the small fluctuations in the spectral parameters obtained while fitting the time-averaged spectra could fit the observed timing features. They applied this scheme to fit the time lag and frms observed for the type-C QPO in GRS 1915+105. Furthermore, \cite{2022MNRAS.514.3285G} employed the same model to analyze the type-C QPO observed in MAXI J1535-571 for a larger set of observations. Following their scheme, \cite{2023MNRAS.525.4515H} also conducted the spectral and timing study for the source H1743-322 using two \textit{Astrosat} observations of the 2016 and 2017 outbursts. In their research, PDS exhibited sharp QPOs and a harmonic component in each set of observations. They extracted the energy-dependent time lag and frms spectra for these components and modeled them using the same scheme introduced by \cite{2020MNRAS.498.2757G}. This scheme has been solely applied to QPOs observed in X-ray binary systems. In this work, we utilize their scheme to explain the timing characteristics of the broad features observed in the black hole binary system GX 339-4 using the simultaneous \textit{AstroSat} and \textit{NICER} observations.

\textit{Astrosat}, the first multi-wavelength astronomical observatory in India, features a set of five instruments, including the Soft X-ray Telescope (SXT), Large Area X-ray Proportional Counter (LAXPC), Cadmium Zinc Telluride Imager (CZTI), Scanning Sky Monitor (SSM), and Ultra-Violet Imaging Telescope (UVIT). The LAXPC instrument provides good timing and spectral capabilities in the 3-80 keV band, making it well-suited for the study of X-ray binaries with rapid time variability. Additionally, the SXT gives medium-resolution spectral capabilities in the 0.3-8 keV energy range, which allows to conduct the broadband spectral studies in conjunction with the LAXPC \citep{2016SPIE.9905E..1DY, 2017JApA...38...30A}. \textit{Astrosat} has provided valuable insights into the spectro-timing properties of several X-ray binaries, including newly discovered transients, with a particular emphasis on black hole X-ray binaries. Several studies have been conducted using \textit{Astrosat} observations to investigate the spectro-timing behavior of BHXB. The Neutron Star Interior Composition Explorer (\textit{NICER}) provides the facility to conduct such a study in the soft X-ray band. The X-ray timing instrument (XTI) of \textit{NICER} is being used to study these timing features, which were primarily unexplored in the soft X-ray range. The combined broadband spectro-timing study of the BHXB using simultaneous \textit{NICER} and \textit{Astrosat} data may offer valuable information about these properties. Using simultaneous \textit{NICER}, Insight-HXMT, and \textit{Astrosat} observations, \cite{2019JHEAp..23...29X} attempted to investigate the broadband spectro-timing properties of the BHXB transient Swift J1658.2-4242. \cite{2021MNRAS.505..713J} conducted the spectro-timing study of X-ray transient MAXI J1346-630 using the broad energy range by incorporating simultaneous \textit{Astrosat} and \textit{NICER} observations. Using a stochastic propagation model \citep{2019MNRAS.486.2964M}, they fitted the energy-dependent time lag and frms.

The low mass black hole X-ray transient GX 339-4 was first discovered in 1973 \citep{1973ApJ...184L..67M}, using the OSO-7 satellite's 1-60 keV MIT X-ray detector. The GX 339-4 generally experiences an outburst every 2–3 yr \citep{tetarenko2016watchdog}. Due to its complicated outburst profile and variety of spectral states, this source is one of the most thoroughly studied sources. GX 339-4 frequently exhibits all known X-ray states during its complete cycle \citep{2006ARA&A..44...49R,2011BASI...39..409B,dunn2010global}. During the initial stage of an outburst, the source enters a hard state, where luminosity can reach exceptionally high. According to \cite{hynes2003dynamical}, it has an orbital period of 1.7 days, and the inclination angle is not well constrained \citep{cowley2002optical,zdziarski2019x}. \cite{zdziarski2019x} reported the distance of the source to be  8-12 kpc with a black hole mass of 4-11 \(\textup{M}_\odot\). Previous research has revealed that the GX 339-4 black hole has a very high spin a* $\sim$0.95 \citep{garcia2015x,parker2016nustar} with strong relativistic reflection signatures in the hard and soft states \citep{miller2004evidence,garcia2015x,liu2022rapidly}. The Power Density Spectrum of GX 339-4 displays a variety of features, ranging from broad to narrow (QPOs). For example, \cite{2023ApJ...958..153L} observed the QPOs during the Hard Intermediate State (HIMS). \cite{2024arXiv240712639A} revealed the presence of a QPO accompanied by a harmonic component. Similarly, \cite{2024arXiv240506090S} detected type-C QPOs along with a harmonic feature. \cite{2024arXiv240610607C} also observed two broadband noise components along with a QPO using AstroSat observations. 
\begin{figure}
	\includegraphics[scale=0.21,angle=0]{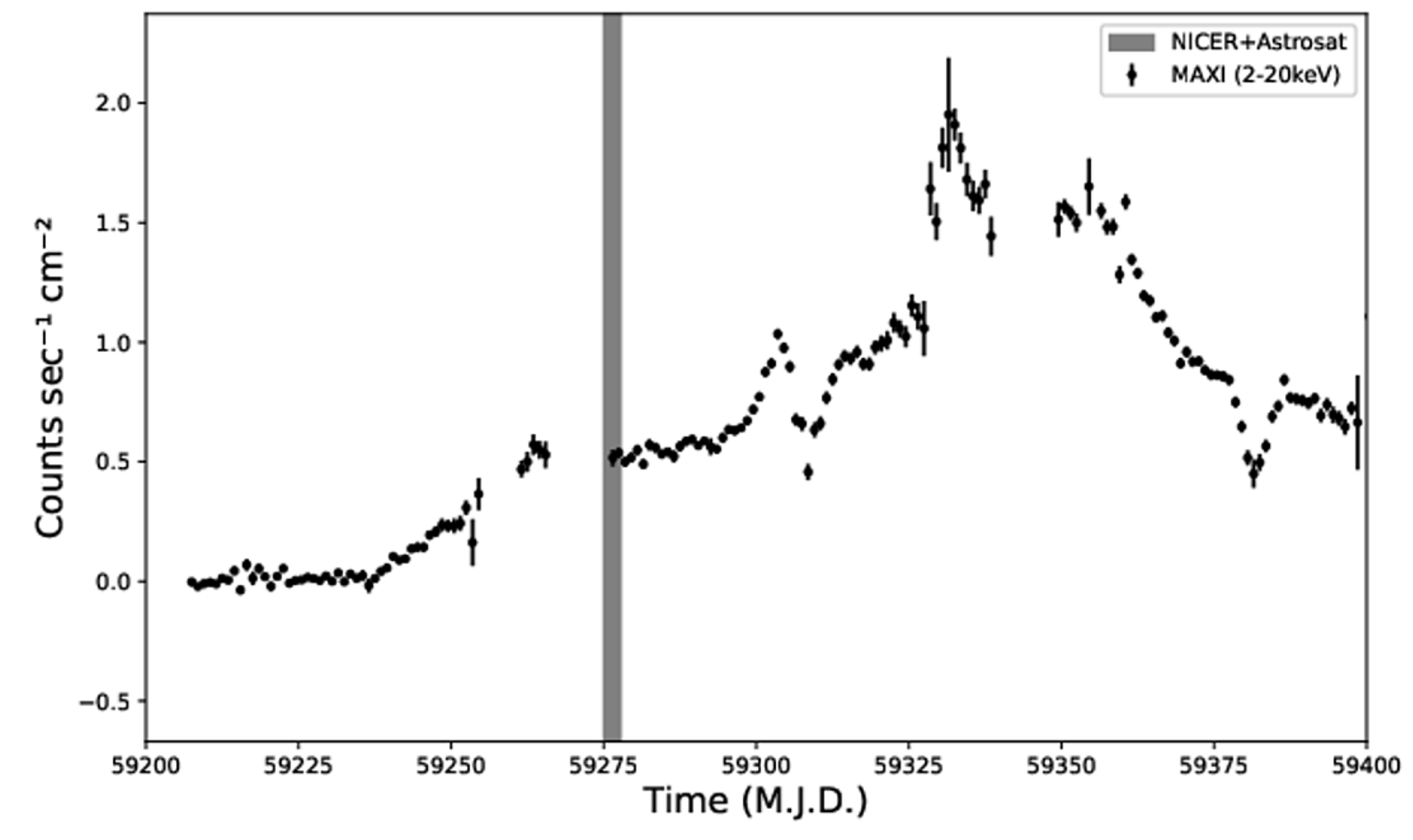}
    \caption{MAXI lightcurve of GX 339-4 in 2-20 keV energy band with the shaded region indicating the simultaneous \textit{NICER} and \textit{Astrosat} observations.}
    \label{fig:example_figure}
\end{figure}

In this work, we present the broadband spectral and temporal characteristics of black hole X-ray transient GX 339-4 during its 2021 outburst using the simultaneous \textit{Astrosat} and \textit{NICER} observations. Our spectral analysis determined the spectral index to be 1.64. Additionally, the power density spectrum displayed low-frequency broadband components. These findings indicate that the source is in a low/hard state \citep{2006ARA&A..44...49R}. The details of the observations and the data reduction process for the \textit{Astrosat} and \textit{NICER} instruments are given in Section 2. In section 3, we present the results obtained through spectral modeling and timing analysis. Finally, we discuss the results and their implications in section 4.

\section{OBSERVATIONS AND DATA REDUCTION} \label{sec:style}

\begin{figure}
\includegraphics[width=\columnwidth, scale=0.4]{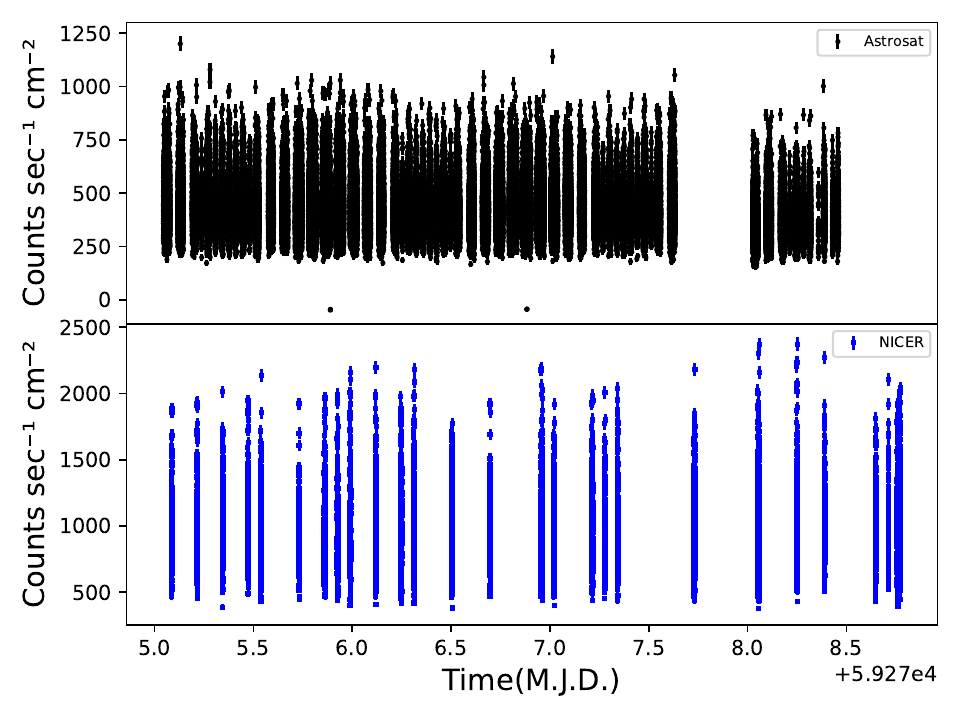}
    \caption{The Upper panel shows the 1.0 sec binned background subtracted lightcurve in 4.0-30.0 keV energy band for LAXPC20 unit and the Lower panel shows the simultaneous \textit{NICER} lightcurve in 0.5-10.0 keV energy band with 1.0 sec binning.}
    \label{fig:example1_figure}
\end{figure}

In this work, we have used Target of Opportunity (ToO) observations of GX 339-4 taken by \textit{AstroSat} from March 2nd to March 5th during its 2021 outburst. We have also used the five \textit{NICER} observations of GX 339-4, which are simultaneous to the \textit{AstroSat}. The details regarding observations are mentioned in Table \ref{Tab:Table 1}. The MAXI light curve in the 2-20 keV energy range for GX 339-4 is shown in Figure \ref{fig:example_figure}. The shaded region marks the simultaneous \textit{Astrosat} and \textit{NICER} observations. Figure \ref{fig:example1_figure} (top panel) depicts the background-subtracted lightcurves of GX 339-4 in a 4–30 keV band as observed by \textit{Astrosat$/$LAXPC20}. In the bottom panel, we show the 0.5–10 keV light curve obtained with simultaneous \textit{NICER} observations. The time bin for both the lightcurves is 1.0 sec.

\subsection{Large Area X-ray Proportional Counter (LAXPC)}

The LAXPC instrument consists of three proportional counters - LAXPC10, LAXPC20, and LAXPC30. The instrument provides a large effective area of approximately 6000 $\mathrm{cm^2}$ and a time resolution of 10 $\mu$s in the energy range of 3.0-80.0 keV \citep{2016SPIE.9905E..1DY,2017ApJS..231...10A,2017JApA...38...30A}. However, among the three units, LAXPC10 and LAXPC30 experienced abnormal behaviors such as low gain and gas leakage. Moreover, LAXPC30 has been non-functional since March 8, 2018. Therefore, in this work, we used only LAXPC20 observations of GX 339-4. We obtained the level 1 data from the ISSDC data archive and extracted level 2 event files using the latest \textbf{LAXPC}\footnote{\url{http://Astrosat-ssc.iucaa.in/laxpcData}} software. The observations were carried out using Event mode, which captures the arrival time and energy of individual incoming photons. We created the Good Time Interval (GTI) file using the LAXPC subroutine "\textit{laxpc$\_$make$\_$stdgti}". Further, we obtained the merged light curve and time-averaged energy spectrum for all observations using the subroutines "\textit{laxpc$\_$make$\_$lightcurve}" and "\textit{laxpc$\_$make$\_$spectra}" respectively.

\subsection{Soft X-ray Telescope (SXT)}
The Soft X-ray Telescope (SXT) operates in the soft X-ray energy range of 0.3-8.0 keV. This instrument offers an effective area of roughly 90 $cm^{2}$ at $\sim$1.5 keV \citep{2017JApA...38...29S}.
We utilized the Photon Counting mode (PC) Level 2 data from SXT. We used the latest \textbf{SXT}\footnote{\url{http://Astrosat-ssc.iucaa.in/sxtData}} software provided by the SXT team for the data reduction. Further, we merged all individual clean event files for each orbit using the SXT$\_$EVENT$\_$MERGER tool. We then extracted out the source events by choosing a circular region of radius 16 arcmin centered on the source using XSELECT (HEASoft version 6.32). We used the latest response matrix file (sxt$\_$pc$\_$mat$\_$g0to12.rmf) and background file (SkyBkg$\_$comb$\_$EL3p5$\_$Cl$\_$Rd16p0$\_$v01.pha) in our analysis. The SXT$\_$ARF$\_$TOOL was used for the vignetting effect, as provided by the \textbf{SXT}\footnote{\url{http://astrosat-ssc.iucaa.in/uploads/threadsPageNew_SXT.html}} team. We utilized the HEASoft package XSELECT to extract the SXT spectrum. Then we grouped the SXT spectrum with the background, Response Matrix File (RMF), and corrected Ancillary Response File (ARF) using the command 'grppha'.
\begin{table}
\centering
 \caption{Log of \textit{Astrosat} and \textit{NICER} Observations of GX 339-4 during 2021 outburst }
 \label{Tab:Table 1}
 \resizebox{0.48\textwidth}{!}{ 
 \begin{tabular}{c c c c}
 \hline\hline\\
 Data& ObsID&Date&Exposure \\
  & &&(in ks)\\
  \hline\hline\\
  \textit{Astrosat}&T03$\_$279T01$\_$9000004218&2021 March 2-4&96.30(LXP)/40.80(SXT)\\ \\
          &T03$\_$280T01$\_$9000004222&2021 March 5&12.02(LXP)/2.07(SXT)\\ \\
          &T03$\_$281T01$\_$9000004224&2021 March 5&1.93(LXP)/0.93(SXT)\\ \\
          &T03$\_$282T01$\_$9000004226&2021 March 5&1.71(LXP)/1.17(SXT)\\ \\
          \hline \\
          \textit{NICER}&3558011201&2021-03-02&3.584\\ \\
&3558011301&2021-03-03&3.371\\ \\
&3558011302&2021-03-04&2.511\\ \\
&3558011401&2021-03-05&1.494\\ \\
&3558010803&2021-03-05&1.358\\ \\

  \hline\hline

 \end{tabular} 
 }
\end{table}

\subsection{NICER}
The Neutron Star Interior Composition Explorer \citep[\textit{NICER}][]{2012SPIE.8443E..13G} is an onboard soft X-ray Telescope that was launched on June 3, 2017. Its X-ray Timing Instrument \citep[\textit{XTI}][]{2016SPIE.9905E..1HG} operates in the energy range of 0.2 to 12 keV and has an energy resolution of 85 keV at 1 keV. \textit{NICER} is also extremely effective at detecting X-ray emissions, with an effective area $>$2000 cm$^{2}$ at 1.5 keV and a good temporal resolution of $\sim 100$ ns.\\

We processed each observation using the {\sc{nicerl2}}\footnote{\url{https://heasarc.gsfc.nasa.gov/docs/nicer/analysis_threads/nicerl2/}} task, which applies the standard calibration process and screening. Additionally, for the five \textit{NICER} observations, we combined the clean event files (output of {\sc{nicerl2}}), auxiliary files (MKF), and unfiltered event (ufa) files. We discarded the events from detectors 14 and 34 using a merged clean event file since they occasionally exhibit increased electronic noise. Further, we used {\sc{nicerl3-spect}}\footnote{\url{https://heasarc.gsfc.nasa.gov/docs/nicer/analysis_threads/nicerl3-spect/}} tool to extract standard spectral products. The grouped spectrum is obtained through the {\sc {nicerl3}} tool, which also incorporates systematic errors and quality flags into the spectrum. {\sc{nicerl3}} uses CALDB version 20221001 to generate auxiliary (ARF), response (RMF), and background files. We used the $3C50$ model in {\sc {nicerl3-spect}} for the background generation.

\section{ANALYSIS AND RESULTS} \label{sec:floats}
\subsection{SPECTRAL ANALYSIS}

We conducted a joint spectral fitting using the merged spectra of LAXPC, SXT and \textit{NICER} instruments in the broad energy range of 0.7-30.0 keV. Due to the uncertainty in the response calibration, a systematic error of $2\%$ is introduced in LAXPC and SXT spectra. We first modeled the broadband spectrum using {\tt Diskbb} and {\tt ThComp} to account for thermal disk emission from the outer truncated disk and Comptonised emission from the inner coronal region, respectively. {\tt Diskbb} model \citep{1984PASJ...36..741M} 
characterizes the disk through two parameters: inner disk temperature (${kT}_{\rm in}$) and disk normalization $(N_{disk})$. We used {\tt ThComp}, a convolution model that serves as a substitute for {\tt nthcomp} to effectively model the non-thermal spectrum emitted by a spherical source of electrons \citep{2020MNRAS.492.5234Z}. It is characterized by four parameters, including the spectral index ($\Gamma$), electron temperature ($kT_{e}$), scattering fraction ($f_{sc}$), and redshift ($z$) of the source. The scattering fraction ($f_{sc}$) represents the proportion of soft seed photons that undergo Comptonization by the hot medium. Among these parameters, {\tt ThComp} allows for the use of either the spectral index or the Thomson optical depth as a fitting parameter \citep{1996MNRAS.283..193Z,zycki19991989,2015MNRAS.448..703W}. Instead of using $\Gamma$ as a fitting parameter, we opted to use Thomson optical depth ($\tau$) as a fitting parameter. This approach was more convenient for predicting the energy-dependent temporal properties such as time lag and frms. To account for the absorption of the incoming photons from the source due to the interstellar medium, we used the XSPEC component {\tt Tbabs} \citep{2000ApJ...542..914W}. Additionally, we modeled the iron emission line at 6.4 keV using the XSPEC model {\tt Gaussian}.\\

\begin{figure*}
\centering

    \includegraphics[scale=0.33,angle=270]{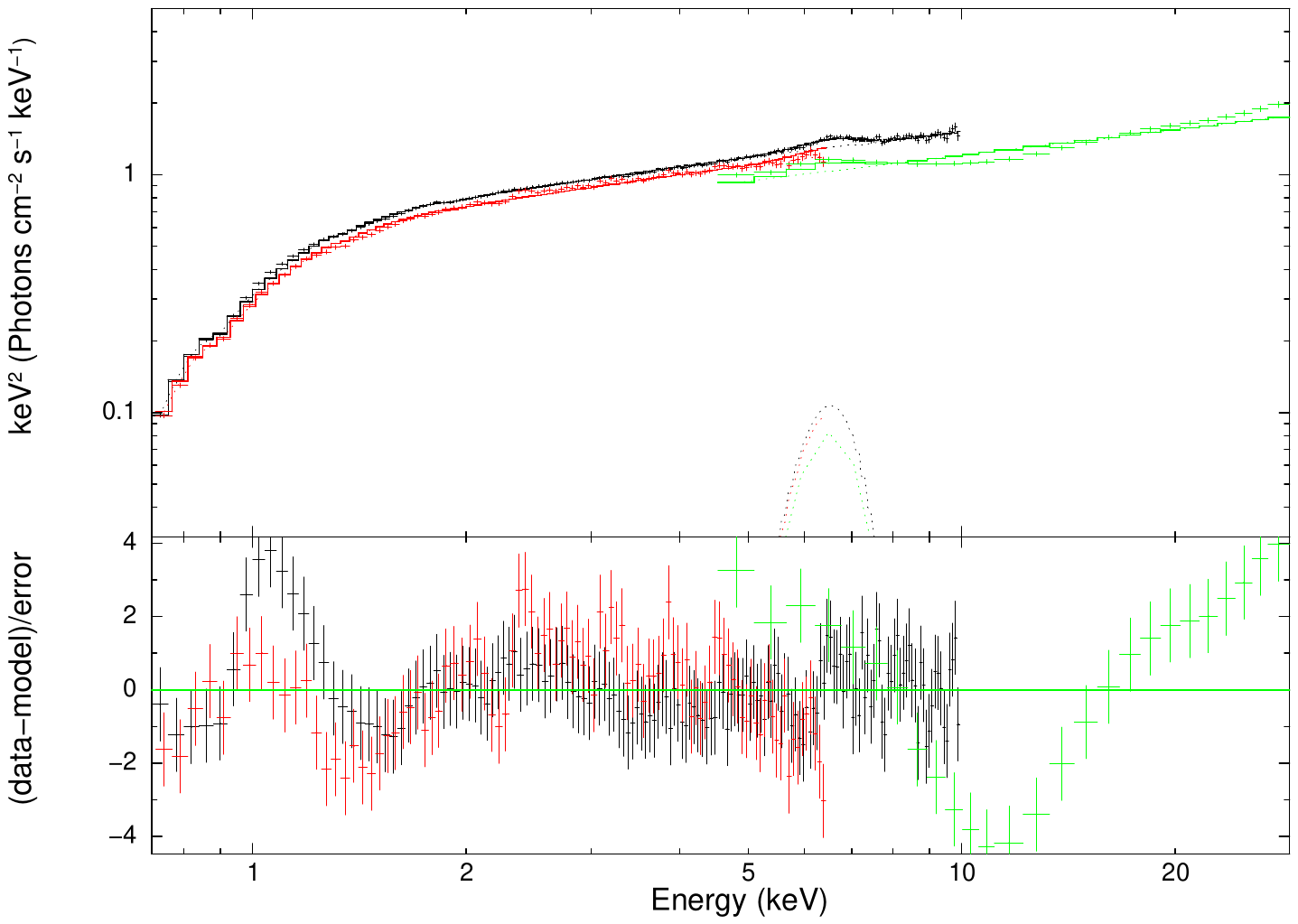}%
    \includegraphics[scale=0.33,angle=270]{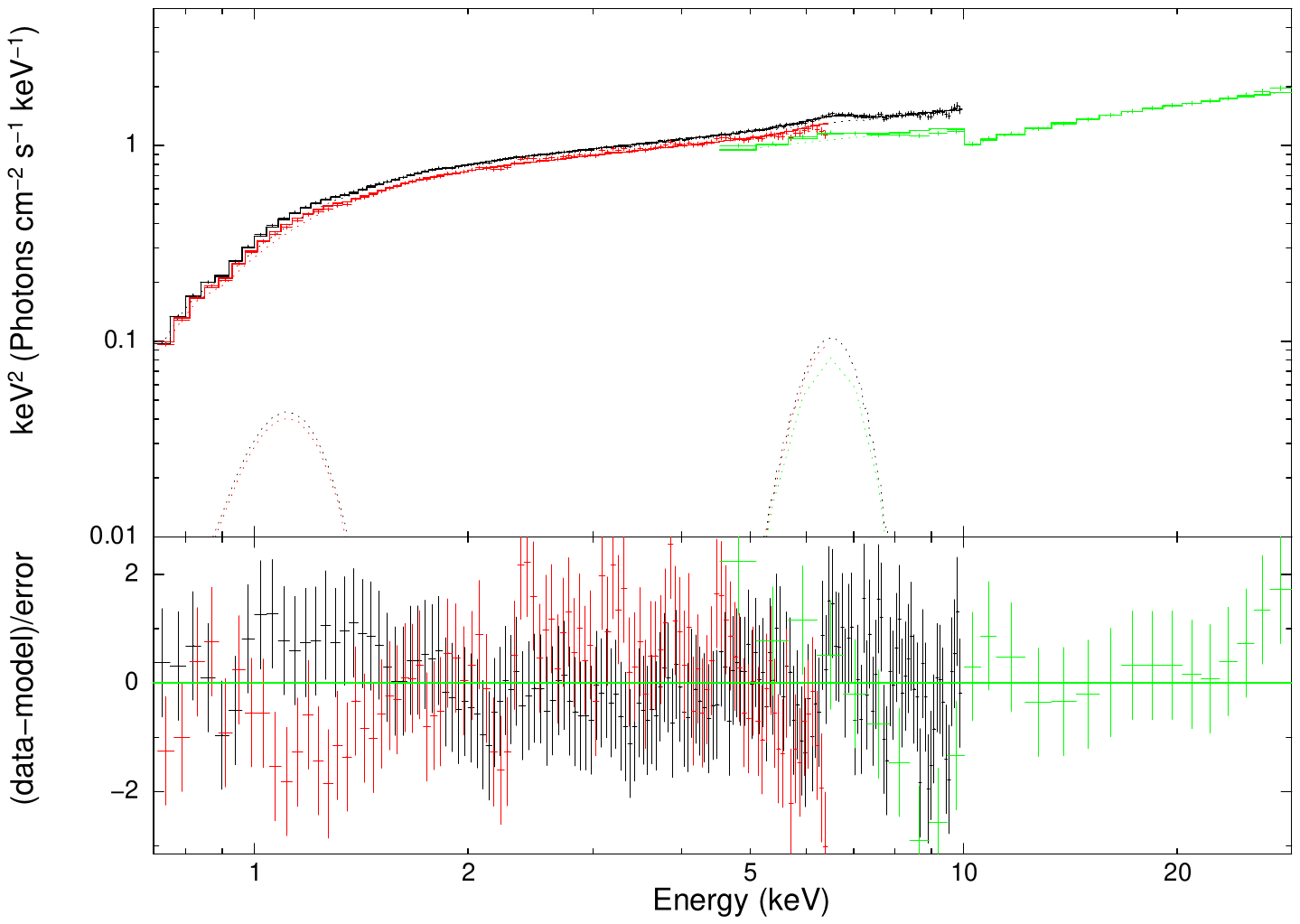} 

\caption{Photon spectrum in 0.7-30.0 keV as observed by SXT (Red), LAXPC20 (Green), and \textit{NICER} (Black). The left upper panel shows the fitting of data with XSPEC Model: {\tt constant*Tbabs*(ThComp*Diskbb+Gaussian)} and the Right upper panel shows the fitting with Model: {\tt constant*Tbabs*(edge*ThComp*Diskbb+Gaussian+Gaussian)}. The bottom panels show residuals.}
\label{fig:spectra_figure}
\end{figure*}

During our analysis, we couldn't constrain the electron temperature ($kT_{e}$) and fixed it to 100 keV. The spectral fitting with the model combination of {\tt constant*Tbabs*(ThComp*Diskbb+Gaussian)} gave a chi-square of 458.98 for 276 degrees of freedom. The best-fit values of spectral parameters are reported in Table \ref{Tab:Table 3}. Figure \ref{fig:spectra_figure} (left panel) shows the simultaneous LAXPC20, SXT and \textit{NICER} spectra fitted using model1 { \tt constant*Tbabs*(ThComp*Diskbb + Gaussian)}.
We observed residuals at $\sim$1.03 keV and $\sim$9.9 keV and thereby included an {\tt edge} model and an additional {\tt Gaussian} to resolve them. We call this model combination {\tt constant*Tbabs*(edge*ThComp*Diskbb+Gaussian} {\tt+Gaussian)} model2, which improved the fitting, resulting in a chi-square value of 230.19 for 271 degrees of freedom. The best-fit parameters for this fitting are listed in Table \ref{Tab:Table 3}. Figure \ref{fig:spectra_figure} (right panel) shows the best-fitted spectrum with the model2 combination. 

\begin{figure*}
\centering
    
    \includegraphics[scale=0.55,angle=0]{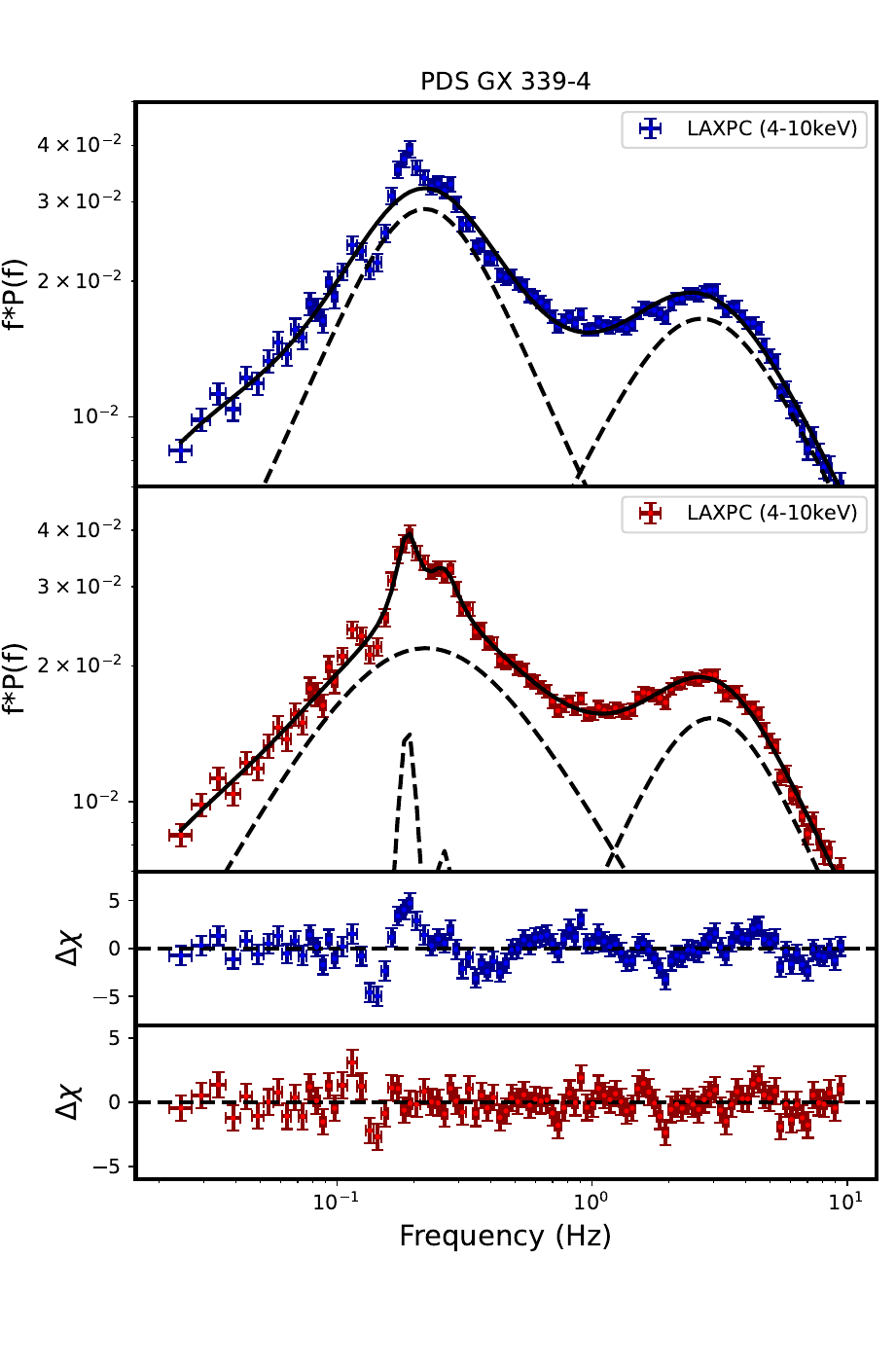}%
    \includegraphics[scale=0.562,angle=0]{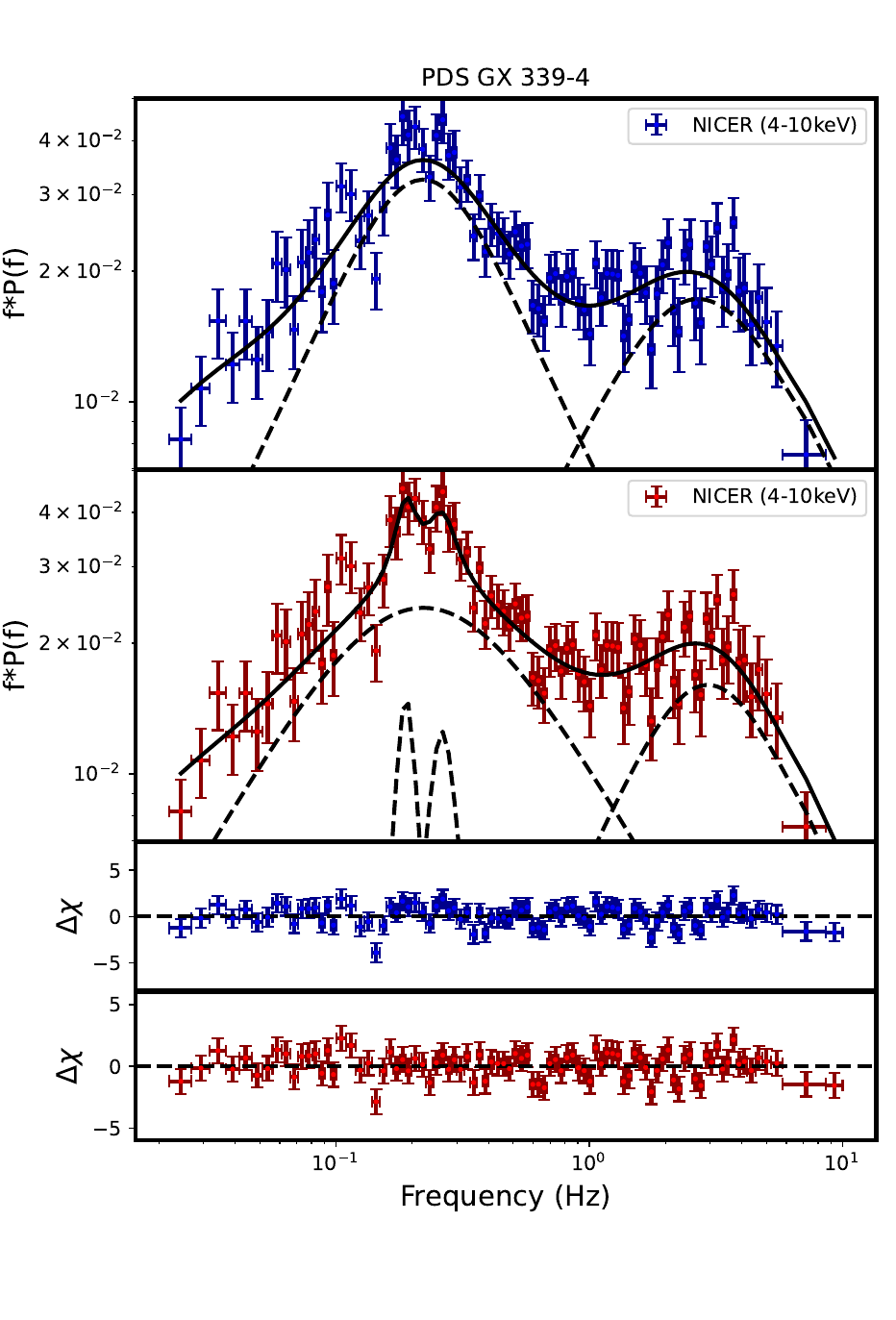}

\caption{The left and right first panels show the PDS of GX 339-4 from \textit{Astrosat/LAXPC20} (4-10 keV) and \textit{NICER} (4-10 keV) respectively as fitted with Lmodel1 along with their residuals in third panels. The left and right second panels show the same PDS but are fitted with Lmodel2 along with their residuals in the fourth panels.}
\label{fig:two_plots}
\end{figure*}

\begin{table}
\centering
 \caption{This Table contains the best-fit values of spectral parameters with model1; {\tt constant*Tbabs*(ThComp*Diskbb+Gaussian)} and model2; {\tt constant*Tbabs*(edge*ThComp*Diskbb+Gaussian+Gaussian)}. The $\dagger$ sign represents the fixed parameters. The error bars are within 90\% confidence level. $N_{H}$ represents the neutral hydrogen column density, measured in units of $ 10^{22}$ atoms $cm^{-2}$. $\tau$ and ${kT}_{\rm e}$ refer to the Thomson optical depth and the temperature of the Comptonized electron, respectively. $f_{sc}$ indicates the covering fraction, while ${kT}_{\rm in}$ denotes the inner disc temperature. $\sigma$ represents the width of the Gaussian line. MaxTau represents the absorption depth.}
 \begin{tabular}{cccc}
 
 \hline\hline\\
 
 Model &Parameter&model1&model2\\ \\
 \hline\hline \\
 $\textit{Tbabs}$&$N_{H}$$ (10^{22} cm^{-2})$&$0.43^{+0.01}_{-0.01}$&$0.36^{+0.01}_{-0.01}$\\ \\
 
 \textit{ThComp}& $\tau$&$1.38^{+0.02}_{-0.02}$&$1.44^{+0.02}_{-0.02}$\\ \\
 
 &${kT}_{\rm e}$ (keV)&$100^{\dagger}$&$100^{\dagger}$\\ \\
 
 &$f_{sc}$ & $0.79^{+0.03}_{-0.04}$& $0.94^{+0.03}_{-0.02}$\\ \\

 \textit{Diskbb}&${kT}_{\rm in}$ (keV)&$0.32^{+0.01}_{-0.01}$&$0.42^{+0.03}_{-0.03}$\\ \\
 
 & $N_{disc}$&$10913^{+3034.77}_{-2154.59}$&$3054.40^{+958.98}_{-784.36}$\\ \\
 
 \textit{Gaussian}&LineE (keV)&$6.4^{\dagger}$&$6.4^{\dagger}$\\ \\
 
 &$\sigma$ (keV)&$0.63^{+0.12}_{-0.10}$&$0.59^{+0.10}_{-0.11}$\\ \\
 
 &$Norm$ $(10^{-3})$&$4.12^{+0.001}_{-0.001}$&$3.76^{+0.001}_{-0.001}$\\ \\
 
  \textit{Gaussian}&LineE (keV)&-&$1.03^{+0.03}_{-0.05}$\\ \\
 
 &$\sigma$ (keV)&-&$0.15^{+0.04}_{-0.03}$\\ \\
 
 &$Norm$ $(10^{-2})$&-&$3.06^{+0.02}_{-0.01}$\\ \\

  $\textit{edge}$&edgeE (keV)&-&$9.94^{+0.08}_{-0.05}$\\ \\
 
 &MaxTau&-&$0.25^{+0.04}_{-0.04}$\\ \\
 \hline
  $\rm \chi^2/ d.o.f$ &&458.98/276&230.19/271\\ 
  
 \hline\hline\\
\label{Tab:Table 3}
 \end{tabular}   
\end{table}

\subsection{TIMING ANALYSIS}
To perform the timing analysis, we barycentered the event files to the barycenter of the Solar System using the "\textit{barycorr}" routine for \textit{NICER} and the "\textit{as1bary}" routine for LAXPC. We have not included the SXT data for the timing analysis due to its insufficient time resolution. \\

The Power Density Spectrum (PDS) provides a detailed description of the distribution of power in the frequency domain, allowing us to investigate the underlying processes driving the variability. To accurately quantify the rapid variations in the X-ray emission of the source, we obtained the PDS for LAXPC20 using subroutine \textit{laxpc$\_$find$\_$freqlag}\footnote{\url{https://www.tifr.res.in/~astrosat_laxpc/LaxpcSoft.html}} in the energy range of 4.0-10.0 keV.
 The PDS was created using the light curve with a time bin size of 0.05 seconds, which corresponds to a Nyquist frequency of 10 Hz. The light curve was divided into segments of time length 204.8 seconds each, and the PDS of these segments were averaged to produce the final PDS. The minimum frequency in the PDS is 0.005 Hz. The PDS is Poisson noise subtracted, assuming a dead time of 42 $\mu$s \citep{yadav2016astrosat}. In the PDS, we limited our analysis to frequencies up to 10 Hz, as no significant variability was detected beyond this value.
The PDS shows two broad features (Hereafter, BF1 and BF2), as shown in the Left panel of Figure \ref{fig:two_plots}. We modeled these broad features using multiple Lorentzian components where each component is characterized by the following equation:- \\
\begin{equation}
L(\nu) = \frac{r^{2} \Delta}{\pi} \frac{1}{(\nu - \nu_{0})^2 + \Delta^{2}}
\label{lor_eqn}
\end{equation}\\
 Here, 'r' represents the integrated frms of each Lorentzian component, over the range from $-\infty$ to $+\infty$, $\nu_{0}$ denotes centroid frequency, and '$\Delta$' denotes its half-width at half-maximum (HWHM). The Lorentzian's characteristic frequency$/$peak frequency, which corresponds to the highest frequency it covers, is expressed as $\nu_{max}$ = $\sqrt{{\nu_{0}}^{2}+ \Delta^{2}}$ \citep{2002ApJ...572..392B}.\\
 
 We fitted the Power Density Spectrum for LAXPC20 using three Lorentzian components (we call this Lmodel1). Specifically, we used two Lorentzian functions to fit the two prominent broad features (hereafter, we call them BF1 and BF2), while a third zero-centered Lorentzian was required to account for the broadband noise component. The best fit $\chi^2$ for this model came out to be 255.17 for 93 d.o.f. Further, we added two more Lorentzian (L4 and L5) to the combination to improve the fitting (We call this as Lmodel2) and obtained a $\chi^2$ of 96.97 for 88 d.o.f. However, we note that L1 and L2 Lorentzians are $24.37 \sigma$ and $57.69 \sigma$ significant, whereas two new Lorentzians, L4 and L5, which improve the fitting are $4.32 \sigma$ and $5.73 \sigma$ significant, respectively. The statistical significance level is computed as the ratio between the best fitting normalization and its $1\sigma$ negative error (see \cite{2014MNRAS.445.4259A}). As the L4 and L5 features are less significant than the L1 and L2 features and overlap with the L1 feature, we chose to focus on the L1 and L2 features only. We obtained the characteristic frequencies as $0.22^{+0.02}_{-0.01}$Hz for BF1 and $2.94^{+0.10}_{-0.12}$Hz for BF2. 
For NICER, we generated the PDS in the energy range of 4.0-10.0 keV with a time resolution of 0.05 seconds, using the subroutine \textit{NICER$\_$find$\_$freqlag} of \textit{NICER$\_$RM$\_$software}\footnote{Developed by Prof. R. Misra, IUCAA}. The light curve was divided into segments of a time length of 204.8 seconds, with a Nyquist frequency of 10 Hz and a minimum frequency of 0.005 Hz. The Poisson noise was subtracted from the PDS, assuming a dead time of nearly zero seconds {\footnote{\url{https://heasarc.gsfc.nasa.gov/docs/nicer/data_analysis/workshops/NICER-Workshop-QA-2021.pdf}}}. Notably, the resulting \textit{NICER} PDS exhibited similar broad features as observed in the LAXPC. As a result, we applied the same Lorentzian fitting that was obtained for LAXPC PDS to model the \textit{NICER} PDS, keeping the LineE and Width parameters fixed to the values obtained by fitting LAXPC PDS and only allowing the Norm to vary for each Lorentzian component. The fitted PDS is represented in the right panel of Figure \ref{fig:two_plots}. The best-fit parameters for LAXPC PDS fitting are listed in Table \ref{tab:lore}.
\\

\renewcommand\arraystretch{1.5}
\begin{table*}
\centering
\caption{Best fit parameter values obtained by fitting the PDS in 4-10 keV for \textit{Astrosat} and \textit{NICER} with three (Lmodel1) and five (Lmodel2) Lorentzian components. The reported errors are within 90$\%$ confidence level. ${\dagger}$sign represents the fixed parameter. Here, $\sigma_1$, $\sigma_2$, $\sigma_3$, $\sigma_4$, and $\sigma_5$ represent the widths, and $N_{L1}$, $N_{L2}$, $N_{L3}$, $N_{L4}$, and $N_{L5}$ represent the normalizations of the Lorentzian components. For broad features, we use the relation $\nu$ = $\sqrt{{\nu_{0}}^{2}+ \Delta^{2}}$ to calculate the characteristic frequency ($\nu$), where $\nu_{0}$ corresponds to LineE and $\Delta$ is the half-width at half-maximum (HWHM=$\sigma/2$).
 } 

\begin{tabular}{ccccccc} \hline \hline
 \textbf{Component}& \textbf{Parameter} &\multicolumn{2}{c}{\textbf{Astrosat}}& \multicolumn{2}{c}{\textbf{NICER}}  &  \\
 \hline
&  & \textbf{Lmodel1}&\textbf{Lmodel2}& \textbf{Lmodel1}&\textbf{Lmodel2}  \\

\hline \hline

\textsc{{L1}} & $LineE$ (Hz) & $0.131^{+0.012}_{-0.011}$ & $0.0^{\dagger}$&$0.131^{\dagger}$& $0.0^{\dagger}$ \\
       (BF1)  & $\sigma_1$ (Hz) & $0.353^{+0.013}_{-0.013}$ & $0.444^{+0.035}_{-0.022}$& $0.353^{\dagger}$ & $0.444^{\dagger}$ \\
               & $N_{L1}~(10^{-2})$ & $6.436^{+0.003}_{-0.003}$ &$6.883^{+0.003}_{-0.005}$& $7.249^{+0.003}_{-0.003}$ & $7.580^{+0.005}_{-0.005}$ \\
               & $fRMS_{L1}~($\%$)$ & $25.37$ & $26.23$&$26.92$&$27.53$\\
               \hline
 \textsc{{L2}} & $LineE$ (Hz) & $1.190^{+0.132}_{-0.140}$ & $1.694^{+0.106}_{-0.114}$& $1.190^{\dagger}$& $1.694^{\dagger}$\\
     (BF2) & $\sigma_2$ (Hz) & $4.813^{+0.098}_{-0.096}$ &$4.798^{+0.121}_{-0.119}$ &$4.813^{\dagger}$&$4.798^{\dagger}$\\
               & $N_{L2}~(10^{-2})$ & $4.164^{+0.001}_{-0.001}$ &$3.470^{+0.001}_{-0.001}$& $4.353^{+0.002}_{-0.002}$& $3.627^{+0.003}_{-0.003}$  \\
            & $fRMS_{L2}~($\%$)$ & $20.40$ & $18.62$& $20.86$& $19.04$ \\
            
               \hline
 \textsc{{L3}} & $LineE$ (Hz) & $0.0^{\dagger}$ &$0.0^{\dagger}$ &$0.0^{\dagger}$&$0.0^{\dagger}$\\
               & $\sigma_3$ (Hz) & $0.062^{+0.017}_{-0.016}$ & $0.049^{+0.036}_{-0.033}$& $0.062^{\dagger}$& $0.049^{\dagger}$\\
               & $N_{L3}~(10^{-2})$ & $1.851^{+0.002}_{-0.002}$ &$1.182^{+0.002}_{-0.002}$ & $2.144^{+0.004}_{-0.004}$& $1.448^{+0.005}_{-0.005}$\\
               & $fRMS_{L3}~($\%$)$ & $13.60$ & $10.87$& $14.64$& $12.03$ \\
              
       \hline  
  \textsc{{L4}}   & $LineE$ (Hz) & - &$0.258^{+0.012}_{-0.017}$& - & $0.258^{\dagger}$\\
               & $\sigma_4$ (Hz) & - &$0.085^{+0.035}_{-0.027}$& - &$0.085^{\dagger}$ \\
               & $N_{L4}~(10^{-2})$ & - &$0.383^{+0.002}_{-0.001}$& - &$0.617^{+0.002}_{-0.002}$ \\ 
               & $fRMS_{L4}~($\%$)$ & - & $6.19$ & - &$7.85$\\
               \hline
     \textsc{{L5}}   & $LineE$ (Hz) & - &$0.187^{+0.004}_{-0.004}$& - &$0.187^{\dagger}$ \\
               & $\sigma_5$ (Hz) & - &$0.043^{+0.014}_{-0.011}$& - &$0.043^{\dagger}$ \\
               & $N_{L5}~(10^{-2})$ & - &$0.511^{+0.002}_{-0.001}$ & - &$0.526^{+0.002}_{-0.002}$\\ 
               & $fRMS_{L5}~($\%$)$ & - & $7.15$& - &$7.25$ \\  
               \hline
                           
\textbf{$\chi^2/dof$}  & & \textbf{255.17/93} & \textbf{96.97/88}& \textbf{109.37/87}& \textbf{90.87/85} \\

\hline 
\label{tab:lore}
\end{tabular}

\end{table*}

\setlength{\tabcolsep}{2.0pt} 
\renewcommand\arraystretch{1.5}
 \begin{table*}
\hspace{-2.0cm}
   \centering
    \caption{Best fit values of parameters obtained by fitting the fractional r.m.s. and time-lag spectra for the observed broad features BF1 (0.22 Hz) and BF2 (2.94 Hz) using Model1, Model2 and Model3. All the error bars are within 90\% confidence level. Here, $|\delta N_{disc}|$, $|\delta kT_{in}|$,  $|\delta \dot{H}|$, and $|\delta f_{sc}|$ represent the variations in disk Normalisation, disk temperature, heating rate, and scattering fraction respectively. $T_{\dot{H}}$ and $T_{f_{sc}}$ represents the time delays of $\delta \dot{H}$ and $\delta (f_{sc})$ with respect to $\delta (N_{disc})$ respectively.}
    \begin{tabular}{cccccccccc}
    \hline \hline
    
         Parameters &\multicolumn{3}{c}{BF1}&\multicolumn{3}{c}{BF2}  &  \\

          \hline
          & M3 & M2 & M1 & M3 & M2 & M1 & \\
          \hline

         $|\delta N_{disc}|$ (\%) &  $44.5^{+9.8}_{-9.8}$ & $25.9^{+0.6}_{-0.6}$ & $30.9^{+0.2}_{-0.2}$ & $35.6^{+8.5}_{-8.2}$ & $16.2^{+0.2}_{-0.2}$ & $20.8^{+0.1}_{-0.1}$ &  \\

          $|\delta kT_{in}|$ (\%) &  $4.8^{+2.5}_{-2.5}$ & $-$ &$- $ & $5.1^{+2.2}_{-2.2}$ & $-$ & $-$ & \\

         $|\delta \dot{H}|$ (\%)  &  $28.0^{+0.9}_{-0.9}$ &  $26.5^{+0.4}_{-0.4}$ & $ 28.5^{+0.4}_{-0.4}$ & $24.4^{+0.7}_{-0.7}$ &  $23.2^{+0.5}_{-0.5}$ & $31.7^{+0.3}_{-0.3}$ &  \\
          
         $|\delta f_{sc}|$ (\%)  & $7.7^{+0.8}_{-0.8}$  & $7.0^{+0.7}_{-0.7}$  & $-$  & $10.5^{+0.6}_{-0.6}$  & $9.5^{+0.4}_{-0.4}$ & $-$ & \\

           $T_{\dot{H}}$ (msec) &  $146.2^{+10.9}_{-10.7} $ & $153.9^{+10.6}_{-10.6}$ & $98.4^{+8.2}_{-8.1}$ & $19.6^{+1.1}_{-1.1} $ & $20.8^{+1.0}_{-1.0}$ &  $16.6^{+0.8}_{-0.9}$ & \\                      
          
          $T_{f_{sc}}$ (msec) &  $311.9^{+50.1}_{-44.6} $ & $341.2^{+53.7}_{-48.9}$ & $-$ & $0$ & $0$ & $-$ &\\           
          \hline
          $\chi^2$/dof    &  \textbf{0.97 (19.51/20)}  & \textbf{1.38 (29.15/21)}  & \textbf{16.87 (388.06/23)}  &\textbf{1.14 (22.91/20)}& \textbf{1.79 (37.67/21)} &\textbf{66.58 (1464.75/22)}&  \\
        \hline    
    \end{tabular}
    \label{tab:5}
\end{table*}
 Next, we generated the fractional r.m.s and time-lag in the 4-30 keV energy range of both features BF1 and BF2 using LAXPC subroutine \textit{laxpc$\_$find$\_$freqlag}. This subroutine uses frequency resolution ($\Delta$f) and the frequency (f) of characteristic features as input parameters to estimate frms and time lag. The subroutine calculates the frms by taking the square root of the value obtained by integrating the power spectra in the frequency range $f-\Delta f$ and $f+\Delta f$ for a particular energy band. Typically, $\Delta$f is set to be equal to or half of the Full Width Half Maximum (FWHM) of the broad frequency for which we want to calculate the time lag. Simultaneously, the subroutine computes the phase lag of the cross-spectrum of two light curves in different energy bands using one as a reference energy band. The time lags are obtained by dividing the phase lag by $2\pi f$ (see \cite{1999ApJ...517..355N} for more details). Likewise, we used \textit{NICER$\_$find$\_$freqlag} to calculate the frms and time-lag for \textit{NICER} observations. For BF1 and BF2, we have calculated the lag energy spectrum in the frequency ranges
of 0.11-0.33 Hz and 1.47-4.41 Hz, respectively. For \textit{Astrosat} and \textit{NICER}, we used the frequency resolution ($\Delta$f) of 0.11 Hz for BF1 and 1.47 Hz for BF2.\\
 
We used 4-5 keV as a reference energy band to calculate the time lag at both frequencies BF1 and BF2 for LAXPC and \textit{NICER} observations in a broad energy range of 0.5-30 keV \footnote{The reference energy band chosen for calculating the time lag was selected to minimize the error bars associated with the time lag.}.  
 We observed that for both \textit{NICER} and \textit{LAXPC}, the time lag is increasing as a function of energy (Figure \ref{fig:lagh1}). In the case of \textit{NICER}, the frms increases for the lower values of energies and, after that, decreases. On the other hand, the frms pattern for \textit{LAXPC} consistently shows a decreasing trend. At BF1 and BF2 frequencies, we observed an offset in the frms of \textit{NICER} and \textit{Astrosat} and thereby multiplied a constant factor with the LAXPC frms values to eliminate this offset. For BF1 and BF2, we used factors of 1.12 and 1.10, respectively, as shown in Figure \ref{fig:lagh1}.

\subsection{MODELLING THE ENERGY-DEPENDENT FRACTIONAL R.M.S. AND TIME-LAGS}

We modeled the observed energy-dependent frms and time-lags of both features BF1 and BF2 using the generic scheme introduced by \cite{2020MNRAS.498.2757G} (hereafter \citetalias{2020MNRAS.498.2757G}). The \citetalias{2020MNRAS.498.2757G} scheme assumes that the time-averaged photon spectrum is mainly composed of {\sc{xspec}} components like {\tt Diskbb} and {\tt nthcomp} to describe the multi-color blackbody and non-thermal Comptonized emissions of the source. The spectral parameters of both these models are then transformed into physical parameters that can better convey the physical and geometrical properties of the disc and the coronal flow. For instance, the parameter electron temperature ($KT_e$) of the component {\tt nthcomp} is converted into coronal heating rate ($\dot{H}$) (as shown in equation 2 of \citetalias{2020MNRAS.498.2757G}) to account for coronal energy output during heating and cooling of the corona. Further, the scheme estimates the first-order variation in the photon spectrum with small fluctuations in the physical parameters using,

\begin{equation}
  \Delta F (E) = \sum_{j=1}^M \frac{\partial F (E)}{\partial \alpha_j} \Delta \alpha_j
  \label{DeltaSE}
\end{equation}\\
\noindent

Where $F(E)$ represents the steady-state spectrum, and $\alpha_j$ are physical parameters. There are a total of $M$ physical parameters. The variations in these parameters, denoted as $\Delta \alpha_j$, can generally be complex numbers. The frms as a function of energy is calculated by $(\frac{1}{\sqrt{2}})
|\Delta F(E)|/F(E)$. Additionally, the phase lag of photons at energy $E$ with respect to a reference energy $E_{ref}$ is defined as the argument of the complex conjugate of ${\Delta F(E_{ref})}$ multiplied by $\Delta F(E)$. In \cite{2022MNRAS.514.3285G} (hereafter, \citetalias{2022MNRAS.514.3285G}), they modified the scheme by including {\tt ThComp} model instead of {\tt nthcomp} to account for hard component and predict the energy-dependent fractional rms and phase-lag. 

We used the updated scheme (\citetalias{2022MNRAS.514.3285G})) to fit the observed frms and lags of broad features BF1 and BF2. In Section 3.1, we fitted the photon spectrum with the model combination of {\tt constant*Tbabs*edge} {\tt(*ThComp*Diskbb+Gaussian+Gaussian)} and obtained the best-fit spectral parameters, namely inner disk temperature (${kT}_{\rm in}$), disk normalization $(N_{disk})$, optical depth ($\tau$), electron temperature ($kT_{e}$), and scattering fraction ($f_{sc}$). Now, we refit the photon spectrum with the model combination {\tt constant*Tbabs*(edge*} {\tt ThCompph*Diskbb+Gaussian+ Gaussian)} where {\tt ThCompph} is a modified version of {\tt ThComp} which uses coronal heating rate ($\dot{H}$) instead of electron temperature during spectral fitting. We kept all the parameters fixed to their values obtained in the former spectral fitting and calculated the best-fit value of the heating rate.

Next, we fit the frms and lag energy spectrum for BF1 and BF2 in {\sc{xspec}} using the same methodology of \citetalias{2020MNRAS.498.2757G} and \citetalias{2022MNRAS.514.3285G}. We used both \textit{NICER} and LAXPC spectra simultaneously to fit in a broad energy band of 0.5-30.0 keV. We started with a simpler model which considers the variations in the disk normalization ($|\delta N_{disc}|$) and heating rate ($|\delta \dot{H}|$), with a phase-lag ($\phi_{\dot{H}}$) between them (with respect to $|\delta N_{disc}|$) (hereafter Model 1). We obtained a poor fit with a high reduced chi-square value of 16.87 for BF1 and 66.58 for BF2. To improve the fit, we introduced variability in the scattering fraction with a phase-lag $\phi_{ f_{sc}}$ (with respect to $|\delta N_{disc}|$) (hereafter, Model2), resulting in reduced chi-square values of 1.38 for BF1 and 1.79 for BF2. However, these fits still did not provide a satisfactory interpretation of the data. Moreover, if we take variation in inner disk temperature instead of variation in scattering fraction, it makes the fit worse with a high chi-square value. 

\begin{figure*}
\centering
    \includegraphics[scale=0.715,angle=0]{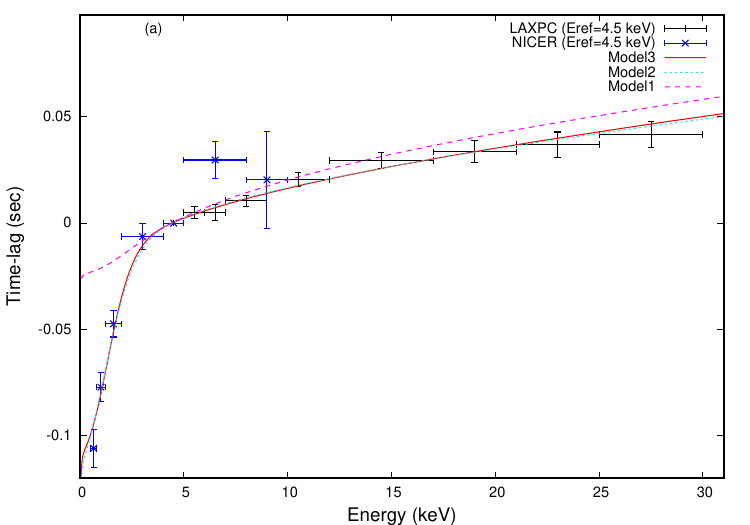}%
     \includegraphics[scale=0.715,angle=0]{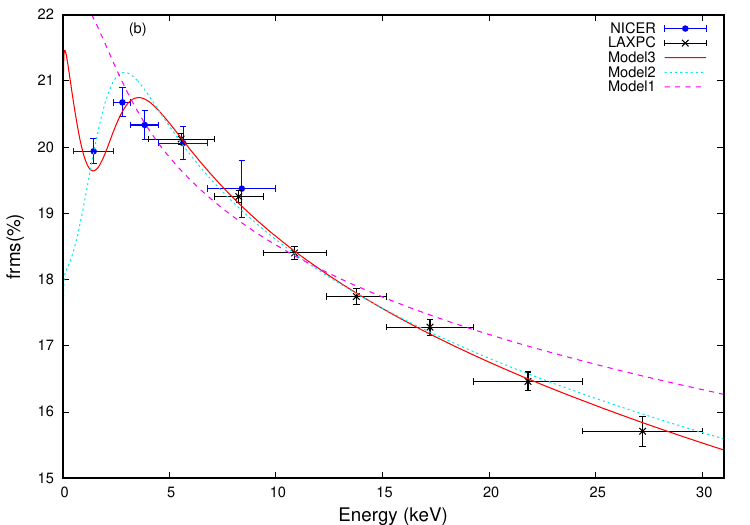}
    \includegraphics[scale=0.715,angle=0]{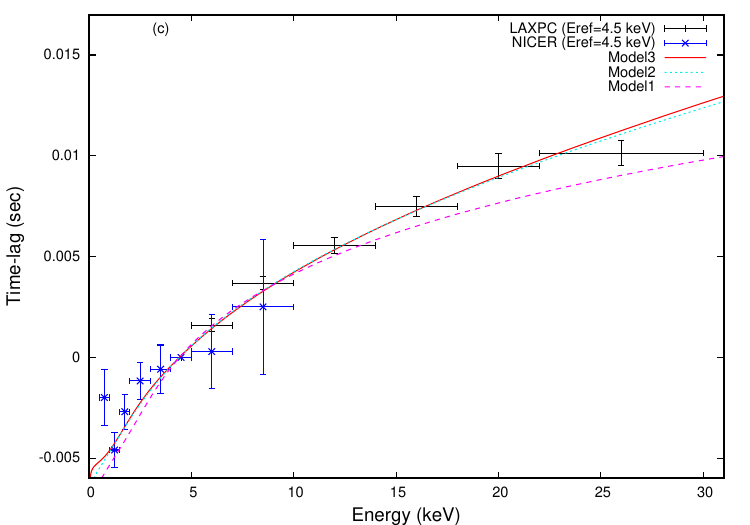}%
        \includegraphics[scale=0.715,angle=0]{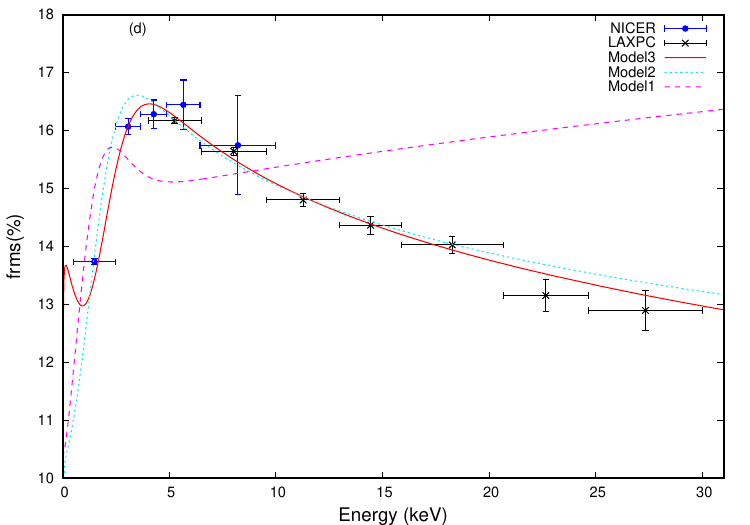}

   \caption{The upper panels show the fitted time lag (a) and frms (b) for BF1 at 0.22 Hz using Model 1, Model 2, and Model 3 for \textit{NICER} (0.5–10.0 keV) and \textit{LAXPC/Astrosat} (4.0–30.0 keV). The lower panels show the fitted time lag (c) and frms (d) for BF2 at 2.94 Hz employing Model 1, Model 2, and Model 3 for \textit{NICER} (0.5–10.0 keV) and \textit{LAXPC/Astrosat} (4.0–30.0 keV). \textbf{Note:}
In the sign convention used here, a positive time lag indicates a hard lag, where the hard energy photons lag behind the soft energy photons at a given Fourier frequency. Conversely, a negative time lag indicates a soft lag, where the soft energy photons lag behind the hard energy photons at a given frequency \citep{2022ApJ...932..113B}.}
    
\label{fig:lagh1}
\end{figure*}

To address this, we further considered Model3, where inner disk temperature varies inversely and without any time delay with respect to $|\delta N_{disc}|$ in addition to variations in the heating rate and scattering fraction with a phase delay of $\phi_{\dot{H}}$ and $\phi_{ f_{sc}}$ with respect to $|\delta N_{disc}|$. This revised model gave a better fit with a chi-square value of 19.51 for 20 degrees of freedom for BF1 and 22.91 for 20 degrees of freedom for BF2. The best-fit parameter values are reported in Table \ref{tab:5}. The fitted time lag and frms spectra are illustrated in Figure \ref{fig:lagh1}.


For BF1, we find that the variations begin with inner disk temperature, which varies the disc normalization inversely without any time delay. These variations further move inwards towards the black hole and cause changes in the heating rate first and then the scattering fraction after a time delay of $\sim$ 146 ms and $\sim$ 312 ms, respectively. However, for BF2, we find that the perturbations begin with inner disc temperature and disc normalization and cause variations in the heating rate as before but vary the scattering fraction with no delay. Here, we observe a time lag of $\sim$ 20 ms between the variations in disc normalization and the heating rate. In Table \ref{tab:5}, note that the time lag between variations in disc normalization and the scattering fraction is zero. When the phase lag corresponding to this time lag is allowed to vary freely during the fitting, it takes the value of $-0.0626^{+0.0655}_{-0.0658}$, with a reduced chi-square of 1.07. However, this best-fit value is quite small and anomalously oscillates between positive and negative values, which is physically implausible. As a result, we fixed this parameter to zero,  as this does not significantly impact the remaining best-fit parameters.

\section{DISCUSSION AND CONCLUSIONS}
We have conducted the joint spectro-timing study of low-mass black hole binary system GX 339-4 during its low/hard spectral state of the 2021 outburst.  In this work, we have used the simultaneous observations taken by \textit{NICER} and \textit{Astrosat}. We find that the time-averaged energy spectrum can be described using a thermal disk, a non-thermal comptonization, and a {\tt Gaussian} reflection component. Additionally, we require an extra {\tt Gaussian} and an {\tt edge} component to resolve the residuals at $\sim$1.03 keV and $\sim$9.9 keV in the energy spectrum. The emission feature around 1 keV is potentially indicative of the Fe-L line, attributed to highly ionized atoms. This is consistent with the findings of \cite{2018ApJ...858L...5L}, who observed a similar feature in the low-mass X-ray binary Serpens X-1 using NICER. Additionally, the absorption feature near $\sim$10 keV is typically attributed to the absorption of accretion disk photons by the highly ionized, blue-shifted disk wind \citep{2023MNRAS.520.4889P}. However, confirming this feature would require a dedicated spectroscopic analysis, which is beyond the scope of the present work. Currently, while modeling the rms and time lag spectra, we consider components such as thermal disc and Comptonization of the spectrum, which dominate the observed continuum. We find that such a model can describe the observed rms and lag spectra and thereby refrain from focussing on these absorption features in this work. 

Further, we have detected two prominent broad features (BF1 and BF2) at 0.22 Hz and 2.94 Hz in the PDS. We generated the time lag and frms for both features and found that both features exhibited hard lags. However, we observe a notable distinction between BF1 and BF2 below $\sim 3$keV for the time lags. Specifically, BF1 displays a rapid decrease in time lag, dropping by approximately $\sim100$ milliseconds. On the other hand, the decrease in time lag for BF2 is relatively small, around $\sim5$ milliseconds. For the frms, both the broad features (BF1 and BF2) show a dip below $\sim3$keV. As we know, the characteristic frequencies of BF1 and BF2 are 0.22 Hz and 2.94 Hz, respectively, suggesting that the variability of BF1 occurs on a longer timescale compared to BF2. This implies that the variability in BF1 originates at a greater distance from the corona within the accretion disc than the variability associated with BF2. As a result, the signal for BF1 takes more time to propagate to the coronal region, potentially explaining the sharp decline in time lag observed for BF1 below 3 keV. This decline suggests that the softer photons generated farther out in the disc are associated with a longer timescale. In contrast, for BF2, the signal is likely to reach the coronal region in a shorter time, which could account for the time lag behavior observed for BF2 below 3 keV. Moreover, \cite{2011MNRAS.414L..60U} also observed a similar decline in hard time lag at lower frequencies within the soft energy band, attributing it to a decrease in the amplitude of disk variability at lower frequencies compared to higher ones. Their lag-energy spectra show that the downturn observed at lower frequencies takes an upturn at higher frequencies, resulting in soft lags. In our study, although we do not observe an upturn due to our frequency range being limited to 2.94 Hz, the time lag behavior is consistent. Therefore, we can also say that a sharp decline below $\sim$3 keV for BF1 (0.22 Hz), which becomes less as we approach higher frequency (BF2), indicates the decreased contribution of disc variability.\\

Such broad features have also been observed for earlier GX 339-4 outbursts. For instance, \cite{2023ApJ...958..153L} observed such features during the 2007 outburst while the source was in the low/hard state. Additionally, \cite{2022MNRAS.510.4040H} conducted a spectro-timing study of transient GX 339-4 during the 2017 and 2019 outbursts, also in the low/hard state, identifying broad features with low-frequency breaks in the PDS. \cite{2019MNRAS.486.2964M} also detected similar broad features in the PDS of Cygnus X-1. \cite{2019MNRAS.486.2964M} modeled the frms and lags for such broad features using a stochastic propagation model based on thermal Comptonization only. Here, we employ a generic technique developed by \citetalias{2020MNRAS.498.2757G}. Earlier, the technique has been used to describe the frms and lags for Type-C QPOs in GRS 1915+105 \citep{2020MNRAS.498.2757G}, MAXI J1535-571 \citep{2022MNRAS.514.3285G}, and H1743-422 \citep{2023MNRAS.525.4515H}. However, in this work, we now use the technique to model the energy-dependent frms and lag for the broad features. We find that the frms and lags at both frequencies can be explained using the correlated variations in the inner disc temperature, scattering fraction, disk normalization, and heating rate with a delay between them. We discover that the dynamical origin of the observed variability is in the disc and then propagates inward to the corona. We have considered a truncated disk geometry in which the total X-ray emission arises from two main components: The soft emission from the geometrically thin and optically thick truncated accretion disk and the hard X-ray emission resulting from the upscattering of soft seed photons of the disc entering the coronal region. So, the soft disk photons act as a source of seed photons for the corona. Subsequently, If there are any variations in the disk parameters, such as accretion rate and inner disc radii, it will affect the coronal parameters, such as heating rate as well. Interestingly, we found that the inner disc temperature and disc normalization vary in anti-correlation with each other for both features. These variations subsequently vary the heating rate and the scattering fraction after some time delays. However, this is true for BF1 only. For BF2, disk parameters and the scattering fraction vary simultaneously and affect the heating rate after a time delay of $\sim$ 20 ms. This points to a difference in how the disc and coronal parameters are varied to generate the observed timing features for both BF1 and BF2. This could indicate different physical phenomena for both features, but we would need more observations of the source to confirm this. 

Moreover, fractional variations of all parameters are of the same order among both features, but the magnitude of time delays is different. We found the time delay of $\sim 146$ msec and $\sim 20$ msec between disc parameter ($|\delta N_{disc}|$) and coronal parameter ($|\delta \dot{H}|$) variations for BF1 and BF2, respectively, which indicates that for a high-frequency feature that originates closer to the black hole, variations take less time to reach the main energy release region. This is in agreement with \cite{2001MNRAS.327..799K}, who found multiple humps in the PDS of Cyg X-1 and interpreted these features as evidence that perturbations originating at specific distances from the compact object dominate over a range of frequencies. They also proposed that various time scales affect the accretion flow at different distances from the compact object, with propagation time scales being comparable to the perturbation time scales. Further, we can also calculate variations in the accretion rate and inner disc radius here using the variations in inner disc temperature and disk normalization through $\delta R_{in} = (1/2)\delta N_{disc}$ and $\delta \dot{M} = 4\delta kT_{in} + 3\delta R_{in}$. Interestingly, we found that the ratio $\delta R_{in} / \delta \dot{M}$ comes out to be $\sim$ 0.25 for both features which implies $R_{in} \propto \dot{M}^{0.25}$. This indicates a positive correlation between the accretion rate and inner disc radii for both broadband features.

Different characteristic time scales are involved over which the structure of the accretion flow evolves \citep{2002apa..book.....F}. At every radius, the dynamical time scale represents the shortest possible time scale in the inner regions of the accretion flow, defined as the inverse of the orbital frequency. Two other important time scales are the thermal time scale, which is linked to restoring thermal equilibrium, and the viscous time scale, which corresponds to how matter diffuses through the disc as a result of viscous torques. The viscous time scale is slower than the thermal and dynamical time scales. Low-frequency variations correspond to longer time scales and cannot originate near the compact object. Low-frequency mass accretion rate variations must originate in the outer regions of the accretion flow and then propagate inward to the main energy release region, where they emerge as fluctuations in the X-ray flux. Considering characteristic time delays observed in this work for broad features, we believe that stochastic perturbations in inner disc temperature (or accretion rate) propagate at sound speeds and produce variations in the coronal parameters \citep{2020ApJ...889L..36M}. \\

\cite{2011MNRAS.414L..60U} conducted the timing studies for GX 339-4 in the low hard state using the XMM-Newton instrument and thereby could analyze the lag spectra in the softer energy bands. Their findings indicate that at low frequencies, the magnitude of the hard lag is greater in the softer energy band than in the harder energy band. Their analysis revealed a significant increase in these lags within the soft energy band. Similarly, our observations indicate a relatively larger hard time lag at lower frequencies for BF1 in the soft energy band, consistent with the previously established relation $\tau \propto \nu^{-0.7}$, where $\tau$ represents the time lag and $\nu$ the frequency. \cite{2011MNRAS.414L..60U} associated these time lags with the viscous propagation of mass accretion rate fluctuations in the disk, suggesting that the variability originates in the disk and propagates inward to the coronal region. However, in this work, we attribute these time lags to the stochastic propagation of fluctuations in the accretion rate at the sound speed. Furthermore, \cite{2011MNRAS.414L..60U} demonstrated that the covariance spectra show a presence of disc blackbody variability at high frequencies, though with a reduced amplitude. Our observations similarly reveal a decrease in fractional rms amplitude at higher frequencies. Specifically, for BF1 (0.22 Hz), the fractional rms amplitude reaches up to approximately 20\%, whereas for BF2 (2.94 Hz), the maximum variation in fractional rms amplitude is around 16\%.\\

In this study, we have conducted the broadband spectral-timing analysis of the black hole X-ray binary candidate GX 339-4 using simultaneous observations from \textit{Astrosat} and \textit{NICER} during the 2021 outburst. The main findings of our study are outlined as follows.
\begin{itemize}
    \item The broadband spectral analysis conducted in the 0.7–30 keV energy range, using simultaneous observations from \textit{Astrosat} and \textit{NICER}, revealed the source to be in the low/hard spectral state. Within this state, spectra are dominated by Comptonization emissions, with a spectral index of approximately 1.64.
  
    \item The Power Density Spectrum obtained from LAXPC revealed the presence of two distinct humps. These humps were identified as broad features and labeled as BF1 and BF2. Similar features were also observed in the \textit{NICER} PDS. The characteristic frequencies of these broad features turned out to be 0.22 Hz for BF1 and 2.94 Hz for BF2.

      \item For the first time, we determined the energy-dependent temporal characteristics, such as fractional rms and time lag for the broad features observed in the PDS of GX 339-4 within the wide energy band of 0.5–30 keV, utilizing data from the \textit{AstroSat/LAXPC} and \textit{NICER} instruments.

    \item We calculated the energy dependence of the fractional rms at frequencies of BF1 and BF2, which slightly increases for the lower energy bands and then decreases at hard energy bands. This decreasing trend implies that the soft X-ray component dominates the variability in the low/hard state, and Comptonized emission is significantly less variable.

    \item We computed the time lag as a function of energy for both broad features (BF1 and BF2) across multiple energy bands for both LAXPC and \textit{NICER} instruments. Notably, both time lags displayed hard lags, indicating that hard photons lag behind soft photons. However, we noticed a sharp decline in the time lag for BF1 below $\sim 3$keV. \\

    \item We fit the energy-dependent frms and time lags using a generic scheme put forward by \citetalias{2020MNRAS.498.2757G}. This model uses the time-averaged spectral parameters of the thermal disk and hard comptonization component obtained from spectral modeling. The model allows variations in the physical spectral parameters and computes the frms and time lag. We allowed variations in four spectral parameters to fit the frms and time lag at frequencies of 0.22 Hz and 2.94 Hz. Following the fitting of the frms and time lag spectra for both features, we observed that variations primarily occurred in the inner disk temperature initially, which then propagated inward, varying the disk normalization, scattering fraction, and heating rate. In both cases (BF1 and BF2), the variation in the inner disk temperature emerged as the primary driver for the observed variability.

    \end{itemize}

\section*{Acknowledgements}
We would like to thank the anonymous reviewer for helpful comments, which have improved the quality of this work. We are grateful to the LAXPC and SXT teams at TIFR, Mumbai, for sharing the data through the ISSDC archive and providing the essential software tools. HT is grateful to Ms. Nazma Husain (Jamia Millia Islamia) and Mr. Suchismito Chattopadhyay (Calcutta University) for their helpful discussions regarding \textit{NICER} data analysis. HT is also thankful to the Inter-University Centre for Astronomy and Astrophysics (IUCAA) for allowing the opportunity to visit and work on this project. This work is based on the results obtained from \textit{Astrosat} and \textit{NICER} missions. This work has made use of data and software from the High Energy Astrophysics Science Archive Research Center (HEASARC), a service of NASA/GSFC's Astrophysics Science Division. We acknowledge the financial support provided by the Department of Space, Govt of India No.DS\_2B-13012(2)/2/2022-Sec.2.

\section*{Data Availability}
The data utilized in this work is publicly available on the Indian Space Science Data Center (ISSDC) website (\url{http://astrobrowse.issdc.gov.in/astro_archive/archive}) for the \textit{Astrosat} mission and the HEASARC database (\url{https://heasarc.gsfc.nasa.gov}) for the \textit{NICER} mission.\\

\bibliographystyle{aasjournal}
\bibliography{example}

\begin{thebibliography}{}
\expandafter\ifx\csname natexlab\endcsname\relax\def\natexlab#1{#1}\fi
\providecommand{\url}[1]{\href{#1}{#1}}

\bibitem[{{Agrawal} {et~al.}(2017){Agrawal}, {Yadav}, {Antia}, {Dedhia}, {Shah}, {Chauhan}, {Manchanda}, {Chitnis}, {Gujar}, {Katoch}, {Kurhade}, {Madhwani}, {Manojkumar}, {Nikam}, {Pandya}, {Parmar}, {Pawar}, {Roy}, {Paul}, {Pahari}, {Misra}, {Ravichandran}, {Anilkumar}, {Joseph}, {Navalgund}, {Pandiyan}, {Sarma}, \& {Subbarao}}]{2017JApA...38...30A}
{Agrawal}, P.~C., {Yadav}, J.~S., {Antia}, H.~M., {et~al.} 2017, Journal of Astrophysics and Astronomy, 38, 30

\bibitem[{{Alam} {et~al.}(2014){Alam}, {Dewangan}, {Belloni}, {Mukherjee}, \& {Jhingan}}]{2014MNRAS.445.4259A}
{Alam}, M.~S., {Dewangan}, G.~C., {Belloni}, T., {Mukherjee}, D., \& {Jhingan}, S. 2014, \mnras, 445, 4259

\bibitem[{{Aneesha} {et~al.}(2024){Aneesha}, {Das}, {Katoch}, \& {Nandi}}]{2024arXiv240712639A}
{Aneesha}, U., {Das}, S., {Katoch}, T.~B., \& {Nandi}, A. 2024, arXiv e-prints, arXiv:2407.12639

\bibitem[{{Antia} {et~al.}(2017){Antia}, {Yadav}, {Agrawal}, {Verdhan Chauhan}, {Manchanda}, {Chitnis}, {Paul}, {Dedhia}, {Shah}, {Gujar}, {Katoch}, {Kurhade}, {Madhwani}, {Manojkumar}, {Nikam}, {Pandya}, {Parmar}, {Pawar}, {Pahari}, {Misra}, {Navalgund}, {Pandiyan}, {Sharma}, \& {Subbarao}}]{2017ApJS..231...10A}
{Antia}, H.~M., {Yadav}, J.~S., {Agrawal}, P.~C., {et~al.} 2017, \apjs, 231, 10

\bibitem[{{Ar{\'e}valo} \& {Uttley}(2006)}]{2006MNRAS.367..801A}
{Ar{\'e}valo}, P., \& {Uttley}, P. 2006, \mnras, 367, 801

\bibitem[{{Baughman} \& {Becker}(2022)}]{2022ApJ...932..113B}
{Baughman}, D.~C., \& {Becker}, P.~A. 2022, \apj, 932, 113

\bibitem[{Bellavita {et~al.}(2022)Bellavita, Garc{\'\i}a, M{\'e}ndez, \& Karpouzas}]{bellavita2022vkompth}
Bellavita, C., Garc{\'\i}a, F., M{\'e}ndez, M., \& Karpouzas, K. 2022, Monthly Notices of the Royal Astronomical Society, 515, 2099

\bibitem[{{Belloni} {et~al.}(2002){Belloni}, {Psaltis}, \& {van der Klis}}]{2002ApJ...572..392B}
{Belloni}, T., {Psaltis}, D., \& {van der Klis}, M. 2002, \apj, 572, 392

\bibitem[{{Belloni} {et~al.}(2011){Belloni}, {Motta}, \& {Mu{\~n}oz-Darias}}]{2011BASI...39..409B}
{Belloni}, T.~M., {Motta}, S.~E., \& {Mu{\~n}oz-Darias}, T. 2011, Bulletin of the Astronomical Society of India, 39, 409

\bibitem[{{Belloni} {et~al.}(2012){Belloni}, {Sanna}, \& {M{\'e}ndez}}]{2012MNRAS.426.1701B}
{Belloni}, T.~M., {Sanna}, A., \& {M{\'e}ndez}, M. 2012, \mnras, 426, 1701

\bibitem[{{Belloni} \& {Stella}(2014)}]{2014SSRv..183...43B}
{Belloni}, T.~M., \& {Stella}, L. 2014, \ssr, 183, 43

\bibitem[{{Bollimpalli} {et~al.}(2020){Bollimpalli}, {Mahmoud}, {Done}, {Fragile}, {Klu{\'z}niak}, {Narayan}, \& {White}}]{2020MNRAS.496.3808B}
{Bollimpalli}, D.~A., {Mahmoud}, R., {Done}, C., {et~al.} 2020, \mnras, 496, 3808

\bibitem[{{Chand} {et~al.}(2024){Chand}, {Dewangan}, {Zdziarski}, {Bhattacharya}, {Mithun}, \& {Vadawale}}]{2024arXiv240610607C}
{Chand}, S., {Dewangan}, G.~C., {Zdziarski}, A.~A., {et~al.} 2024, arXiv e-prints, arXiv:2406.10607

\bibitem[{Cowley {et~al.}(2002)Cowley, Schmidtke, Hutchings, \& Crampton}]{cowley2002optical}
Cowley, A., Schmidtke, P., Hutchings, J., \& Crampton, D. 2002, The Astronomical Journal, 123, 1741

\bibitem[{Dunn {et~al.}(2010)Dunn, Fender, K{\"o}rding, Belloni, \& Cabanac}]{dunn2010global}
Dunn, R., Fender, R., K{\"o}rding, E., Belloni, T., \& Cabanac, C. 2010, Monthly Notices of the Royal Astronomical Society, 403, 61

\bibitem[{Frank {et~al.}(2002)Frank, King, \& Raine}]{2002apa..book.....F}
Frank, J., King, A., \& Raine, D. 2002, Accretion Power in Astrophysics, 3rd edn. (Cambridge University Press)

\bibitem[{Garc{\'\i}a {et~al.}(2015)Garc{\'\i}a, Steiner, McClintock, Remillard, Grinberg, \& Dauser}]{garcia2015x}
Garc{\'\i}a, J.~A., Steiner, J.~F., McClintock, J.~E., {et~al.} 2015, The Astrophysical Journal, 813, 84

\bibitem[{{Garg} {et~al.}(2020){Garg}, {Misra}, \& {Sen}}]{2020MNRAS.498.2757G}
{Garg}, A., {Misra}, R., \& {Sen}, S. 2020, \mnras, 498, 2757

\bibitem[{{Garg} {et~al.}(2022){Garg}, {Misra}, \& {Sen}}]{2022MNRAS.514.3285G}
---. 2022, \mnras, 514, 3285

\bibitem[{{Gendreau} {et~al.}(2012){Gendreau}, {Arzoumanian}, \& {Okajima}}]{2012SPIE.8443E..13G}
{Gendreau}, K.~C., {Arzoumanian}, Z., \& {Okajima}, T. 2012, in Society of Photo-Optical Instrumentation Engineers (SPIE) Conference Series, Vol. 8443, Space Telescopes and Instrumentation 2012: Ultraviolet to Gamma Ray, ed. T.~{Takahashi}, S.~S. {Murray}, \& J.-W.~A. {den Herder}, 844313

\bibitem[{{Gendreau} {et~al.}(2016){Gendreau}, {Arzoumanian}, {Adkins}, {Albert}, {Anders}, {Aylward}, {Baker}, {Balsamo}, {Bamford}, {Benegalrao}, {Berry}, {Bhalwani}, {Black}, {Blaurock}, {Bronke}, {Brown}, {Budinoff}, {Cantwell}, {Cazeau}, {Chen}, {Clement}, {Colangelo}, {Coleman}, {Coopersmith}, {Dehaven}, {Doty}, {Egan}, {Enoto}, {Fan}, {Ferro}, {Foster}, {Galassi}, {Gallo}, {Green}, {Grosh}, {Ha}, {Hasouneh}, {Heefner}, {Hestnes}, {Hoge}, {Jacobs}, {J{\o}rgensen}, {Kaiser}, {Kellogg}, {Kenyon}, {Koenecke}, {Kozon}, {LaMarr}, {Lambertson}, {Larson}, {Lentine}, {Lewis}, {Lilly}, {Liu}, {Malonis}, {Manthripragada}, {Markwardt}, {Matonak}, {Mcginnis}, {Miller}, {Mitchell}, {Mitchell}, {Mohammed}, {Monroe}, {Montt de Garcia}, {Mul{\'e}}, {Nagao}, {Ngo}, {Norris}, {Norwood}, {Novotka}, {Okajima}, {Olsen}, {Onyeachu}, {Orosco}, {Peterson}, {Pevear}, {Pham}, {Pollard}, {Pope}, {Powers}, {Powers}, {Price}, {Prigozhin}, {Ramirez}, {Reid}, {Remillard}, {Rogstad}, {Rosecrans}, {Rowe}, {Sager}, {Sanders},
  {Savadkin}, {Saylor}, {Schaeffer}, {Schweiss}, {Semper}, {Serlemitsos}, {Shackelford}, {Soong}, {Struebel}, {Vezie}, {Villasenor}, {Winternitz}, {Wofford}, {Wright}, {Yang}, \& {Yu}}]{2016SPIE.9905E..1HG}
{Gendreau}, K.~C., {Arzoumanian}, Z., {Adkins}, P.~W., {et~al.} 2016, in Society of Photo-Optical Instrumentation Engineers (SPIE) Conference Series, Vol. 9905, Space Telescopes and Instrumentation 2016: Ultraviolet to Gamma Ray, ed. J.-W.~A. {den Herder}, T.~{Takahashi}, \& M.~{Bautz}, 99051H

\bibitem[{{Husain} {et~al.}(2023){Husain}, {Garg}, {Misra}, \& {Sen}}]{2023MNRAS.525.4515H}
{Husain}, N., {Garg}, A., {Misra}, R., \& {Sen}, S. 2023, \mnras, 525, 4515

\bibitem[{{Husain} {et~al.}(2022){Husain}, {Misra}, \& {Sen}}]{2022MNRAS.510.4040H}
{Husain}, N., {Misra}, R., \& {Sen}, S. 2022, \mnras, 510, 4040

\bibitem[{Hynes {et~al.}(2003)Hynes, Steeghs, Casares, Charles, \& O'Brien}]{hynes2003dynamical}
Hynes, R.~I., Steeghs, D., Casares, J., Charles, P., \& O'Brien, K. 2003, The Astrophysical Journal, 583, L95

\bibitem[{{Ingram} \& {Motta}(2019)}]{2019NewAR..8501524I}
{Ingram}, A.~R., \& {Motta}, S.~E. 2019, \nar, 85, 101524

\bibitem[{{Jahoda} {et~al.}(2006){Jahoda}, {Markwardt}, {Radeva}, {Rots}, {Stark}, {Swank}, {Strohmayer}, \& {Zhang}}]{2006ApJS..163..401J}
{Jahoda}, K., {Markwardt}, C.~B., {Radeva}, Y., {et~al.} 2006, \apjs, 163, 401

\bibitem[{{Jithesh} {et~al.}(2019){Jithesh}, {Maqbool}, {Misra}, {T}, {Mall}, \& {James}}]{2019ApJ...887..101J}
{Jithesh}, V., {Maqbool}, B., {Misra}, R., {et~al.} 2019, \apj, 887, 101

\bibitem[{{Jithesh} {et~al.}(2021){Jithesh}, {Misra}, {Maqbool}, \& {Mall}}]{2021MNRAS.505..713J}
{Jithesh}, V., {Misra}, R., {Maqbool}, B., \& {Mall}, G. 2021, \mnras, 505, 713

\bibitem[{Karpouzas {et~al.}(2020)Karpouzas, M{\'e}ndez, Ribeiro, Altamirano, Blaes, \& Garc{\'\i}a}]{karpouzas2020comptonizing}
Karpouzas, K., M{\'e}ndez, M., Ribeiro, E.~M., {et~al.} 2020, Monthly Notices of the Royal Astronomical Society, 492, 1399

\bibitem[{{Kotov} {et~al.}(2001){Kotov}, {Churazov}, \& {Gilfanov}}]{2001MNRAS.327..799K}
{Kotov}, O., {Churazov}, E., \& {Gilfanov}, M. 2001, \mnras, 327, 799

\bibitem[{{Kumar} \& {Misra}(2014)}]{2014MNRAS.445.2818K}
{Kumar}, N., \& {Misra}, R. 2014, \mnras, 445, 2818

\bibitem[{Liu {et~al.}(2022)Liu, Jiang, Zhang, Bambi, Ji, Kong, \& Zhang}]{liu2022rapidly}
Liu, H., Jiang, J., Zhang, Z., {et~al.} 2022, Monthly Notices of the Royal Astronomical Society, 513, 4308

\bibitem[{{Lucchini} {et~al.}(2023){Lucchini}, {Ten Have}, {Wang}, {Homan}, {Kara}, {Adegoke}, {Connors}, {Dauser}, {Garcia}, {Mastroserio}, {Ingram}, {van der Klis}, {K{\"o}nig}, {Lewin}, {Mallick}, {Nathan}, {O'Neill}, {Panagiotou}, {Piotrowska}, \& {Uttley}}]{2023ApJ...958..153L}
{Lucchini}, M., {Ten Have}, M., {Wang}, J., {et~al.} 2023, \apj, 958, 153

\bibitem[{{Ludlam} {et~al.}(2018){Ludlam}, {Miller}, {Arzoumanian}, {Bult}, {Cackett}, {Chakrabarty}, {Dauser}, {Enoto}, {Fabian}, {Garc{\'\i}a}, {Gendreau}, {Guillot}, {Homan}, {Jaisawal}, {Keek}, {La Marr}, {Malacaria}, {Markwardt}, {Steiner}, \& {Strohmayer}}]{2018ApJ...858L...5L}
{Ludlam}, R.~M., {Miller}, J.~M., {Arzoumanian}, Z., {et~al.} 2018, \apjl, 858, L5

\bibitem[{{Lyubarskii}(1997)}]{1997MNRAS.292..679L}
{Lyubarskii}, Y.~E. 1997, \mnras, 292, 679

\bibitem[{{Maqbool} {et~al.}(2019){Maqbool}, {Mudambi}, {Misra}, {Yadav}, {Gudennavar}, {Bubbly}, {Rao}, {Jogadand}, {Patil}, {Bhattacharyya}, \& {Singh}}]{2019MNRAS.486.2964M}
{Maqbool}, B., {Mudambi}, S.~P., {Misra}, R., {et~al.} 2019, \mnras, 486, 2964

\bibitem[{{Markert} {et~al.}(1973){Markert}, {Canizares}, {Clark}, {Lewin}, {Schnopper}, \& {Sprott}}]{1973ApJ...184L..67M}
{Markert}, T.~H., {Canizares}, C.~R., {Clark}, G.~W., {et~al.} 1973, \apjl, 184, L67

\bibitem[{{M{\'e}ndez} {et~al.}(2013){M{\'e}ndez}, {Altamirano}, {Belloni}, \& {Sanna}}]{2013MNRAS.435.2132M}
{M{\'e}ndez}, M., {Altamirano}, D., {Belloni}, T., \& {Sanna}, A. 2013, \mnras, 435, 2132

\bibitem[{M{\'e}ndez \& Belloni(2021)}]{Mendez2021}
M{\'e}ndez, M., \& Belloni, T.~M. 2021, High-Frequency Variability in Neutron-Star Low-Mass X-ray Binaries (Berlin, Heidelberg: Springer Berlin Heidelberg), 263--331.
\newblock \url{https://doi.org/10.1007/978-3-662-62110-3_6}

\bibitem[{{M{\'e}ndez} {et~al.}(2022){M{\'e}ndez}, {Karpouzas}, {Garc{\'\i}a}, {Zhang}, {Zhang}, {Belloni}, \& {Altamirano}}]{2022NatAs...6..577M}
{M{\'e}ndez}, M., {Karpouzas}, K., {Garc{\'\i}a}, F., {et~al.} 2022, Nature Astronomy, 6, 577

\bibitem[{Miller {et~al.}(2004)Miller, Fabian, Reynolds, Nowak, Homan, Freyberg, Ehle, Belloni, Wijnands, Van~der Klis, {et~al.}}]{miller2004evidence}
Miller, J.~M., Fabian, A., Reynolds, C., {et~al.} 2004, The Astrophysical Journal, 606, L131

\bibitem[{{Misra} {et~al.}(2020){Misra}, {Rawat}, {Yadav}, \& {Jain}}]{2020ApJ...889L..36M}
{Misra}, R., {Rawat}, D., {Yadav}, J.~S., \& {Jain}, P. 2020, \apjl, 889, L36

\bibitem[{{Mitsuda} {et~al.}(1984){Mitsuda}, {Inoue}, {Koyama}, {Makishima}, {Matsuoka}, {Ogawara}, {Shibazaki}, {Suzuki}, {Tanaka}, \& {Hirano}}]{1984PASJ...36..741M}
{Mitsuda}, K., {Inoue}, H., {Koyama}, K., {et~al.} 1984, \pasj, 36, 741

\bibitem[{{Motta} {et~al.}(2022){Motta}, {Belloni}, {Stella}, {Pappas}, {Casares}, {Mu{\~n}oz-Darias}, {Torres}, \& {Yanes-Rizo}}]{2022MNRAS.517.1469M}
{Motta}, S.~E., {Belloni}, T., {Stella}, L., {et~al.} 2022, \mnras, 517, 1469

\bibitem[{{Mudambi} {et~al.}(2020){Mudambi}, {Maqbool}, {Misra}, {Hebbar}, {Yadav}, {Gudennavar}, \& {S.~G.}}]{2020ApJ...889L..17M}
{Mudambi}, S.~P., {Maqbool}, B., {Misra}, R., {et~al.} 2020, \apjl, 889, L17

\bibitem[{{Nowak}(2000)}]{2000MNRAS.318..361N}
{Nowak}, M.~A. 2000, \mnras, 318, 361

\bibitem[{{Nowak} {et~al.}(1999){Nowak}, {Wilms}, \& {Dove}}]{1999ApJ...517..355N}
{Nowak}, M.~A., {Wilms}, J., \& {Dove}, J.~B. 1999, \apj, 517, 355

\bibitem[{Parker {et~al.}(2016)Parker, Tomsick, Kennea, Miller, Harrison, Barret, Boggs, Christensen, Craig, Fabian, {et~al.}}]{parker2016nustar}
Parker, M., Tomsick, J., Kennea, J., {et~al.} 2016, The Astrophysical Journal Letters, 821, L6

\bibitem[{{Prabhakar} {et~al.}(2023){Prabhakar}, {Mandal}, {Bhuvana}, \& {Nandi}}]{2023MNRAS.520.4889P}
{Prabhakar}, G., {Mandal}, S., {Bhuvana}, G.~R., \& {Nandi}, A. 2023, \mnras, 520, 4889

\bibitem[{{Rawat} {et~al.}(2023){Rawat}, {M{\'e}ndez}, {Garc{\'\i}a}, {Altamirano}, {Karpouzas}, {Zhang}, {Alabarta}, {Belloni}, {Jain}, \& {Bellavita}}]{2023MNRAS.520..113R}
{Rawat}, D., {M{\'e}ndez}, M., {Garc{\'\i}a}, F., {et~al.} 2023, \mnras, 520, 113

\bibitem[{{Remillard} \& {McClintock}(2006)}]{2006ARA&A..44...49R}
{Remillard}, R.~A., \& {McClintock}, J.~E. 2006, \araa, 44, 49

\bibitem[{{Schnittman} {et~al.}(2006){Schnittman}, {Homan}, \& {Miller}}]{2006ApJ...642..420S}
{Schnittman}, J.~D., {Homan}, J., \& {Miller}, J.~M. 2006, \apj, 642, 420

\bibitem[{{Shakura} \& {Sunyaev}(1976)}]{1976MNRAS.175..613S}
{Shakura}, N.~I., \& {Sunyaev}, R.~A. 1976, \mnras, 175, 613

\bibitem[{{Shapiro} {et~al.}(1976){Shapiro}, {Lightman}, \& {Eardley}}]{1976ApJ...204..187S}
{Shapiro}, S.~L., {Lightman}, A.~P., \& {Eardley}, D.~M. 1976, \apj, 204, 187

\bibitem[{{Shyam Prakash V.} {et~al.}(2024){Shyam Prakash V.}, {Ramadevi M.}, \& {Agrawal}}]{2024arXiv240506090S}
{Shyam Prakash V.}, P., {Ramadevi M.}, C., \& {Agrawal}, V.~K. 2024, arXiv e-prints, arXiv:2405.06090

\bibitem[{{Singh} {et~al.}(2017){Singh}, {Stewart}, {Westergaard}, {Bhattacharayya}, {Chandra}, {Chitnis}, {Dewangan}, {Kothare}, {Mirza}, {Mukerjee}, {Navalkar}, {Shah}, {Abbey}, {Beardmore}, {Kotak}, {Kamble}, {Vishwakarama}, {Pathare}, {Risbud}, {Koyande}, {Stevenson}, {Bicknell}, {Crawford}, {Hansford}, {Peters}, {Sykes}, {Agarwal}, {Sebastian}, {Rajarajan}, {Nagesh}, {Narendra}, {Ramesh}, {Rai}, {Navalgund}, {Sarma}, {Pandiyan}, {Subbarao}, {Gupta}, {Thakkar}, {Singh}, \& {Bajpai}}]{2017JApA...38...29S}
{Singh}, K.~P., {Stewart}, G.~C., {Westergaard}, N.~J., {et~al.} 2017, Journal of Astrophysics and Astronomy, 38, 29

\bibitem[{Stella \& Vietri(1997)}]{stella1997lense}
Stella, L., \& Vietri, M. 1997, The Astrophysical Journal, 492, L59

\bibitem[{Stella {et~al.}(1999)Stella, Vietri, \& Morsink}]{stella1999correlations}
Stella, L., Vietri, M., \& Morsink, S.~M. 1999, The Astrophysical Journal, 524, L63

\bibitem[{Tetarenko {et~al.}(2016)Tetarenko, Sivakoff, Heinke, \& Gladstone}]{tetarenko2016watchdog}
Tetarenko, B., Sivakoff, G., Heinke, C., \& Gladstone, J. 2016, The Astrophysical Journal Supplement Series, 222, 15

\bibitem[{{Uttley} {et~al.}(2014){Uttley}, {Cackett}, {Fabian}, {Kara}, \& {Wilkins}}]{2014A&ARv..22...72U}
{Uttley}, P., {Cackett}, E.~M., {Fabian}, A.~C., {Kara}, E., \& {Wilkins}, D.~R. 2014, \aapr, 22, 72

\bibitem[{{Uttley} {et~al.}(2011){Uttley}, {Wilkinson}, {Cassatella}, {Wilms}, {Pottschmidt}, {Hanke}, \& {B{\"o}ck}}]{2011MNRAS.414L..60U}
{Uttley}, P., {Wilkinson}, T., {Cassatella}, P., {et~al.} 2011, \mnras, 414, L60

\bibitem[{{van der Klis}(2005)}]{2005AIPC..797..345V}
{van der Klis}, M. 2005, in American Institute of Physics Conference Series, Vol. 797, Interacting Binaries: Accretion, Evolution, and Outcomes, ed. L.~{Burderi}, L.~A. {Antonelli}, F.~{D'Antona}, T.~{di Salvo}, G.~L. {Israel}, L.~{Piersanti}, A.~{Tornamb{\`e}}, \& O.~{Straniero}, 345--358

\bibitem[{{Veledina}(2016)}]{2016ApJ...832..181V}
{Veledina}, A. 2016, \apj, 832, 181

\bibitem[{{Wilkins} \& {Gallo}(2015)}]{2015MNRAS.448..703W}
{Wilkins}, D.~R., \& {Gallo}, L.~C. 2015, \mnras, 448, 703

\bibitem[{{Wilms} {et~al.}(2000){Wilms}, {Allen}, \& {McCray}}]{2000ApJ...542..914W}
{Wilms}, J., {Allen}, A., \& {McCray}, R. 2000, \apj, 542, 914

\bibitem[{{Xiao} {et~al.}(2019){Xiao}, {Ji}, {Staubert}, {Ge}, {Zhang}, {Zhang}, {Santangelo}, {Ducci}, {Liao}, {Guo}, {Li}, {Zhang}, {Qu}, {Lu}, {Li}, {Song}, {Xu}, {Bu}, {Cai}, {Cao}, {Chang}, {Chen}, {Chen}, {Chen}, {Chen}, {Chen}, {Chen}, {Cui}, {Cui}, {Deng}, {Dong}, {Du}, {Fu}, {Gao}, {Gao}, {Gao}, {Gu}, {Guan}, {Gungor}, {Guo}, {Han}, {Huang}, {Huo}, {Jia}, {Jiang}, {Jiang}, {Jin}, {Kong}, {Li}, {Li}, {Li}, {Li}, {Li}, {Li}, {Li}, {Li}, {Li}, {Liang}, {Liu}, {Liu}, {Liu}, {Liu}, {Liu}, {Lu}, {Lu}, {Luo}, {Luo}, {Ma}, {Meng}, {Nang}, {Nie}, {Ou}, {Sai}, {Song}, {Sun}, {Tan}, {Tao}, {Tuo}, {Wang}, {Wang}, {Wang}, {Wang}, {Wang}, {Wen}, {Wu}, {Wu}, {Wu}, {Xiong}, {Yang}, {Yang}, {Yang}, {Yang}, {Yin}, {Yin}, {Zhang}, {Zhang}, {Zhang}, {Zhang}, {Zhang}, {Zhang}, {Zhang}, {Zhang}, {Zhang}, {Zhang}, {Zhang}, {Zhang}, {Zhang}, {Zhang}, {Zhang}, {Zhang}, {Zhao}, {Zhao}, {Zheng}, {Zhou}, {Zhu}, \& {Zhu}}]{2019JHEAp..23...29X}
{Xiao}, G.~C., {Ji}, L., {Staubert}, R., {et~al.} 2019, Journal of High Energy Astrophysics, 23, 29

\bibitem[{Yadav {et~al.}(2016)Yadav, Misra, Chauhan, Agrawal, Antia, Pahari, Dedhia, Katoch, Madhwani, Manchanda, {et~al.}}]{yadav2016astrosat}
Yadav, J., Misra, R., Chauhan, J.~V., {et~al.} 2016, The Astrophysical Journal, 833, 27

\bibitem[{{Yadav} {et~al.}(2016){Yadav}, {Agrawal}, {Antia}, {Chauhan}, {Dedhia}, {Katoch}, {Madhwani}, {Manchanda}, {Misra}, {Pahari}, {Paul}, \& {Shah}}]{2016SPIE.9905E..1DY}
{Yadav}, J.~S., {Agrawal}, P.~C., {Antia}, H.~M., {et~al.} 2016, in Society of Photo-Optical Instrumentation Engineers (SPIE) Conference Series, Vol. 9905, Space Telescopes and Instrumentation 2016: Ultraviolet to Gamma Ray, ed. J.-W.~A. {den Herder}, T.~{Takahashi}, \& M.~{Bautz}, 99051D

\bibitem[{{Zdziarski} {et~al.}(1996){Zdziarski}, {Johnson}, \& {Magdziarz}}]{1996MNRAS.283..193Z}
{Zdziarski}, A.~A., {Johnson}, W.~N., \& {Magdziarz}, P. 1996, \mnras, 283, 193

\bibitem[{{Zdziarski} {et~al.}(2020){Zdziarski}, {Szanecki}, {Poutanen}, {Gierli{\'n}ski}, \& {Biernacki}}]{2020MNRAS.492.5234Z}
{Zdziarski}, A.~A., {Szanecki}, M., {Poutanen}, J., {Gierli{\'n}ski}, M., \& {Biernacki}, P. 2020, \mnras, 492, 5234

\bibitem[{Zdziarski {et~al.}(2019)Zdziarski, Zi{\'o}{\l}kowski, \& Miko{\l}ajewska}]{zdziarski2019x}
Zdziarski, A.~A., Zi{\'o}{\l}kowski, J., \& Miko{\l}ajewska, J. 2019, Monthly Notices of the Royal Astronomical Society, 488, 1026

\bibitem[{{Zhang} {et~al.}(2022){Zhang}, {M{\'e}ndez}, {Garc{\'\i}a}, {Karpouzas}, {Zhang}, {Liu}, {Belloni}, \& {Altamirano}}]{2022MNRAS.514.2891Z}
{Zhang}, Y., {M{\'e}ndez}, M., {Garc{\'\i}a}, F., {et~al.} 2022, \mnras, 514, 2891

\bibitem[{{\.Z}ycki {et~al.}(1999){\.Z}ycki, Done, \& Smith}]{zycki19991989}
{\.Z}ycki, P.~T., Done, C., \& Smith, D.~A. 1999, Monthly Notices of the Royal Astronomical Society, 309, 561

\end{thebibliography}



\end{document}